\documentclass[useAMS,usenatbib,]{mn2e}
\usepackage{amssymb}
\usepackage{times}
\usepackage{graphicx}
\usepackage{subfigure}
\usepackage{epsfig}
\usepackage{mathrsfs,amssymb}
\usepackage{amsmath}
\usepackage{ulem}
\usepackage{color}

\newcommand{\psec}{\ensuremath{\, {\rm s}^{-1}}}
\newcommand{\yr}{\ensuremath{{\rm yr}}}
\newcommand{\cm}{\ensuremath{{\rm cm}}}
\newcommand{\m}{\ensuremath{{\rm m}}}
\newcommand{\km}{\ensuremath{{\rm km}}}
\newcommand{\pc}{\ensuremath{{\rm pc}}}
\newcommand{\Mpc}{\ensuremath{\, {\rm Mpc}}}
\newcommand{\K}{\ensuremath{{\rm K}}}
\newcommand{\psr}{\ensuremath{{\rm sr}^{-1}}}
\newcommand{\Hz}{\ensuremath{\, {\rm Hz}}}
\newcommand{\kHz}{\ensuremath{\, {\rm kHz}}}
\newcommand{\MHz}{\ensuremath{\, {\rm MHz}}}
\newcommand{\erg}{\ensuremath{{\rm erg}}}
\newcommand{\eV}{\ensuremath{{\rm eV}}}
\newcommand{\keV}{\ensuremath{\, {\rm keV}}}
\newcommand{\Jy}{\ensuremath{{\rm Jy}}}

\title[The earliest galaxies seen in 21 cm line]{The earliest galaxies seen in 21 cm line absorption}
\author[Y. Xu, A. Ferrara, \& X. Chen]{Yidong Xu$^{1,2,4}$\thanks{E-mail: xuyd@vega.bac.pku.edu.cn},
Andrea Ferrara$^{3}$, and Xuelei Chen$^{4,5}$
 \\ $^{1}$Department of Astronomy, School of Physics, Peking
University, Beijing 100871, China
 \\ $^{2}$SISSA/International School for Advanced Studies, Via Beirut 4, 34014 Trieste, Italy
 \\ $^{3}$Scuola Normale Superiore, Piazza dei Cavalieri 7, 56126 Pisa, Italy
 \\ $^{4}$National Astronomical Observatories, Chinese Academy of Sciences,
Beijing 100012, China
\\ $^{5}$Center for High Energy Physics, Peking University, Beijing 100871, China }

\begin{document}

\maketitle

\begin{abstract}
We investigate the 21 cm absorption lines produced by non-linear
structures during the early stage of reionization, i.e. the starless
minihalos and the dwarf galaxies. After a detailed modelling of
their properties, with particular attention to the coupling physics,
we determine their 21 cm absorption line profiles. The infalling gas
velocity around minihalos/dwarf galaxies strongly affects the line
shape, and with the low spin temperatures outside the virial radii
of the systems, gives rise to horn-like line profiles. The optical
depth of a dwarf galaxy is reduced 
for lines of sight penetrating through its HII region, and
especially, a large HII region created by a dwarf galaxy with higher
stellar mass and/or a top-heavy initial mass function results in an
optical depth trough rather than an absorption line. We compute
synthetic spectra of 21 cm forest for both high redshift quasars and
radio afterglows of gamma ray bursts (GRBs). Even with the planned
SKA, radio afterglows of most if not all GRBs would still be too dim
to be the background sources for high resolution (1 kHz)
observations, but absorption lines can be easily detected towards a
high-$z$ quasar. Broadband observation against GRB afterglows can
also be used to reveal the evolving 21 cm signal from both minihalos
and dwarf galaxies if there was no X-ray background or it was
extremely weak, but it becomes difficult if an early X-ray
background existed. Hence the 21 cm absorption could be a powerful
probe of the presence/intensity of the X-ray background and the
thermal history of the early universe.

\end{abstract}
\begin{keywords}
line: profiles -- galaxies: dwarf -- galaxies: high-redshift -- 
cosmology: theory.
\end{keywords}

\section{INTRODUCTION}
The formation of the earliest galaxies and the cosmic reionization
are among the milestones in the history of the universe. As the
first stars form in the earliest non-linear structures, they
illuminate the ambient intergalactic medium (IGM) and start the
reionization process of hydrogen. Based on an instantaneous
reionization model, the polarization data of cosmic microwave
background (CMB) constrain the redshift of reionization to be
$z_{\rm reion}\approx10.5$ \citep{Larson10}, while the Gunn-Peterson
troughs \citep{GP65} shown in quasar (QSO) absorption spectra
suggest that the reionization of hydrogen was very nearly complete
by $z \approx 6$ (e.g. \citealt{Fan06}). However, at present we are
still unable to observe the earliest galaxies directly, and our
current understanding on the reionization process is only based on
theoretical models (e.g. \citealt{Furlanetto04}) and simulations
(e.g. \citealt{Trac08}).

The most promising probe of the cosmic reionization is the
redshifted 21 cm transition of HI which is directly related to the
neutral component of the IGM (see e.g. \citealt{FOB06} for a
review). Unlike the Ly$\alpha$ resonance line, the 21 cm line could
hardly be saturated because of its extremely small Einstein
coefficient ($A_{\rm 10} = 2.85\times 10^{-15}\psec$), so it traces
well the reionization history, especially during the early stages.
Using the CMB as the background radio source, 21 cm tomography could
map out the three dimensional structure of the emission or
absorption of 21 cm photons by the IGM (e.g.
\citealt{Madau97,Tozzi00}). However, it requires the deviation of
the spin temperature from the CMB temperature and may not be
feasible for certain epochs; it could not resolve structures smaller
than $\sim 1 \Mpc$; and it has a number of observational challenges
\citep{FOB06}. Complementary to the 21 cm tomography, the 21 cm
forest observation detects absorption lines of intervening
structures against high redshift radio sources
\citep{Carilli02,Furlanetto02,Furlanetto06}, and it is immune to
most of the above difficulties encountered by the tomography
observation. As it is very sensitive to gas temperature
\citep{Xu09}, it provides a useful tool to constrain the X-ray
heating in the early universe. And also, the 21 cm forest is a
promising probe that could possibly detect high redshift minihalos,
which are important for determining the mean clumping factor of the
IGM and putting limits on small scale structure formation. Here we
focus on the absorption experiments and investigate the 21 cm
absorption lines produced by the non-linear structures during the
epoch of reionization.

The forest observation relies on the availability of luminous radio
sources beyond the epoch of reionization. One possibility is the
high redshift quasars which have been observed up to $z=6.43$
\citep{Willott07}, and a radio-loud quasar at $z=6.12$ was
discovered by \citet{McGreer06}. Using high-z quasars or radio
galaxies as backgrounds, \citet{Carilli02} and \citet{Xu09} have
examined the possibility of detecting 21 cm absorption by the
neutral IGM based on simulations. Another possible option is the
radio afterglows of high-$z$ gamma ray bursts (GRBs). It is believed
that some of the long duration bursts are produced by the explosions
of massive stars. The first stars are thought to be likely very
massive, and may produce bright GRBs which would be detectable up to
a redshift as high as 60 \citep{Naoz07}. Recently, GRB 090423 was
discovered at $z=8.1$ (\citealt{GRB8.1}, while \citealt{GRB8.26}
reported its redshift to be $z=8.26$), establishing the new redshift
record of observation for all objects except for the CMB. The radio
afterglow of a GRB can be observed out to a very late time when the
outflow becomes sub-relativistic (e.g. \citealt{Pihlstrom07}), and
considering the cosmic time dilation, it offers us an adequate
integration time. One possible problem with the GRB radio afterglow
is that at very low frequency, synchrotron self-absorption may
become important which may reduce the radio flux. Nonetheless, some
GRB radio afterglows may be sufficiently bright at the relevant
frequency ranges.

In order to plan such observations, we need to know what kind of
signals will be produced by the early structures, and to
understand the physics behind the expected signals. In this paper,
we provide a detailed modeling of the 21 cm absorption lines produced by
minihalos and dwarf galaxies during the epoch of reionization, and
explore the physical origins of the line profiles. We generate
synthetic spectra of both quasars and GRB afterglows, on top of
which the 21 cm absorption lines are superposed. Projecting the capability of
future instruments, we discuss the prospects of detecting the 21 cm
signals from these non-linear objects in the early universe.

This paper is arranged as follows. In Section \ref{model}, we
describe the physical model involved with the 21 cm absorptions,
including the halo model for high redshifts, the star burst
criterion, a possible X-ray background, the physical processes
taking place in minihalos and dwarf galaxies respectively, and the
Ly$\alpha$ background produced by these early galaxies. In Section
\ref{profiles}, we show the spin temperatures and 21 cm line
profiles of minihalos and dwarf galaxies for various parameters.
Statistical results including line number density, theoretical
spectrum, and the equivalent width (EW) distribution are given in
Section \ref{theoSpec}. Then, in Section \ref{obsSpec} we study the
feasibility of the 21 cm observation by making mock spectra of both
quasar and radio afterglow of GRB. Finally, we summarize and discuss
our results in Section \ref{discuss}.

Throughout this paper, we adopt the cosmological parameters from
WMAP5 measurements combined with SN and BAO data: $\Omega_b =
0.0462$, $\Omega_c = 0.233$, $\Omega_\Lambda = 0.721$, $H_{\rm 0} =
70.1 \km \psec \Mpc^{-1}$, $\sigma_{\rm 8} = 0.817$, and $n_{\rm s}
= 0.96$ \citep{WMAP5}.

\section{THE MODEL}\label{model}
We start with a description of various aspects of physics involved
with determining the absorption spectrum of 21 cm lines. This
includes the high redshift number density of dark matter halos, gas
distribution inside and around halos, the criterion of star
formation, the X-ray and Ly$\alpha$ background, and a detailed
modeling of the physical properties of starless minihalos and
galaxies, respectively. Among these properties, the ionization
state, the gas temperature distribution, and the Ly$\alpha$ photon
density are especially important for determining the strength and
line profiles of the 21 cm absorption by minihalos/dwarf galaxies.

\subsection{The Halo Model}
In order to model the halo number density at high redshift, we use
the Sheth-Tormen halo mass function, which is based on an
ellipsoidal model for perturbation collapse and fits well the
simulation results. The comoving number density of halos at redshift
$z$ with mass in the interval $(M,M+dM)$, can be written as
\citep{ST99}
\begin{equation}\label{Eq.haloMF}
n(M,z)\, dM \,=\, F_{\rm ST}(\sigma,z)\, \frac{\bar\rho_0}{M}\,
\frac{d\ln \sigma^{-1}}{dM}\, dM,
\end{equation}
where
\begin{equation}
F_{\rm ST}(\sigma,z) = A\, \sqrt{\frac{2a}{\pi}} \left[1 + \left(
\frac{ \sigma^2}{a\, \delta^2_{sc}} \right)^p\, \right]
\frac{\delta_{sc}}{\sigma}\, \exp\left[-\, \frac{a\,
\delta^2_{sc}}{2\, \sigma^2} \right].
\end{equation}
Here $\bar\rho_0$ is the cosmic mean density of the total matter
today, $\sigma = \sigma (M)$ is the r.m.s. of a Gaussian density
field smoothed on a mass scale $M$ with a spherical top-hat filter
of radius $R$, where $R$ is equivalent to $M$ in a fixed cosmology
as $R=(3M/4\pi \bar\rho_0)^{1/3}$, and $\delta_{sc} = 1.686/D(z)$ is
the critical overdensity required for spherical collapse at redshift
$z$, extrapolated to the present time using the linear theory, where
$D(z)$ is the linear growth factor. The correction factors $a =
0.707$, $p = 0.3$, and $A = 0.3222$ were introduced as appropriate
for ellipsoidal collapse \citep{ST02,SMT01}.

The 21 cm signal depends on the gas content of minihalos or
galaxies, and the gas fraction in halos will be suppressed by the
heating processes during the reionization. Therefore, we set the
lower limit of the halo mass to be the characteristic mass $M_{\rm
C}$ at which halos on average could only retain half of their
baryons. The characteristic mass depends on the halo merger history
and the thermal evolution of the universe
\citep{Gnedin00,Okamoto08}. At high redshift that we are considering
($z>6$), however, the filtering mass $M_{\rm F}$ provides a good fit
to the characteristic mass $M_{\rm C}$ \citep{Okamoto08}. Including
the global heating process by an X-ray background (see section
\ref{OpticalDepthWithXRB} in the following), we compute the thermal
evolution of the universe and find that this mass scale is $\sim
10^6 M_\odot$ beyond redshift 7 for an early X-ray background not
higher than $20\%$ of the intensity today. For high redshifts of
interest, it is not very sensitive to the uncertain intensity of the
X-ray background because of the delayed response of the gas density
distribution to the change in the gas temperature. However, if the
early X-ray background was as high as today, or even higher, it
would have moderate effect in raising the Jeans mass and the
filtering mass (see the right panel of Fig.\ref{Fig.TIGM_xi_MFMJ}).
So we take the halo mass range of $[10^6 M_\odot, 10^{10} M_\odot]$
for an X-ray background not higher than $20\%$ of today's value, but
also consider its effect on the minimum halo mass for higher
intensities of the X-ray background (see section \ref{theoSpec}).
The range covers the characteristic halo mass and most of the
galaxies that are responsible for the reionization \citep{CF07}.

In the following, we use the NFW density profile for dark matter
distribution inside of the virial radius $r_{\rm vir}$ of a halo
\citep{NFW97}. In general, the key parameter in the NFW profile, the
concentration parameter $c$, depends on the halo mass as well as its
redshift. Unfortunately, the concentration parameter found from low
redshift simulations \citep{CS02} is not directly applicable to the
epoch of reionization.  Thanks to the resimulation technique,
\cite{Gao05} have simulated an example of the first halos from very
high redshift with extremely high resolution, and its density
profile is derived in each resimulation at the final time. Here we
make use of their results. Assuming that $c$ is inversely
proportional to $(1+z)$, as indicated by low redshift results, we
make a 4th-order polynomial fit to the simulated points
in the logarithmic space of halo mass. 


For the distribution of gas within the dark matter halo, we assume
that the gas is in hydrostatic equilibrium with the dark matter, and
have a spherical distribution. We expect gas in the high-$z$ (and in
general low-mass) galaxies have a rounder distribution rather than
settling into a disc because: (a) their circular velocities are
comparable to the sound speed, making the gas more rounded than
flat; (b) at such high redshifts, there may not be enough time for
the gas to settle in a smooth disk. With the NFW profile of dark
matter distribution, one can derive the gas density distribution
analytically \citep{Makino98}:
\begin{equation}\label{Eq.rho_g}
\ln \rho_{\rm g}(r) \,=\, \ln \rho_{\rm g0} \,-\, \frac{\mu\, m_p}
{2\, k_{\rm B}\, T_{\rm vir}}\, \left[\, v_e^2(0) - v_e^2(r)\,
\right],
\end{equation}
where $\rho_{\rm g0}$ is the central gas density, $\mu$ is the mean
molecular weight of the gas depending on the ionization state, $m_p$
is the proton mass, $k_{\rm B}$ is the Boltzmann constant, and
$T_{\rm vir}$ is the virial temperature of the halo. Here $v_e(r)$
denotes the gas escape velocity at radius $r$, which can be written
as
\begin{equation}
v_e^2(r) = 2 \int_r^{\infty} \frac{GM(r^{\prime})}{r^{\prime 2}}\,
dr^{\prime} \,=\, 2\, V_c^2\, \frac{\displaystyle
F(c\,x)+\frac{c\,x}{1+c\,x}}{x\,F(c)},
\end{equation}
where $V_c^2 \equiv GM/\,r_{\rm vir}$ is the circular velocity at
the virial radius, $x \equiv r/\,r_{\rm vir}$, $c$ is the halo
concentration, and $F(c) = \ln(1+c) - c/(1+c)$. The maximum escape velocity
is reached at the center of the halo, $v_e^2(0) = 2\, V_c^2\,
c/F(c)$. In Eq.(\ref{Eq.rho_g}), the central density $\rho_{\rm g0}$
is determined by the condition that the total baryonic mass fraction
within the virial radius is equal to $\Omega_b/\Omega_m$,
which gives
\begin{equation}
\frac{\rho_{\rm g0}}{\bar\rho_{\rm m}} \,=\, \frac{(\Delta_c/3)\,
c^3\, (\Omega_b/\Omega_m)\, e^A} {\int_0^c (1+t)^{A/t}\, t^2 \, dt},
\end{equation}
where $\bar\rho_{\rm m} = \bar\rho_{\rm m} (z)$ is the mean matter
density of the universe at redshift $z$, $A \equiv 2\,c/F(c)$, and
$\Delta_c$ is the mean density of a virialized halo with respect to the
cosmic mean value $\bar\rho_{\rm m}$ \citep{BN98}: $\Delta_c \,=\,
18\,\pi^2 + 82\left(\Omega_m^z-1\right) -
39\left(\Omega_m^z-1\right)^2$, where $\Omega_m^z = \Omega_m(1+z)^3
/\, [\Omega_m(1+z)^3 + \Omega_\Lambda]$.

To determine the gas distribution outside $r_{\rm vir}$, one has to
keep in mind that the gas slowly falls into the halo because of the
gravitational force of the halo, so the gas has a peculiar (i.e.
infalling) velocity which could be important for the 21 cm line.
Also, due to the large cross-section at large radii, there is a high
probability for the line of sight to go through these external parts
of the halos, and this infalling gas has to be included in the
calculation. \citet{Barkana04} has derived a model for the expected
profiles of infalling matter around virialized halos based on the
extended Press-Schechter (EPS) formalism \citep{Bond91}. This
``Infall Model''\footnote{Public code for this ``Infall Model'' is
available at
http://wise-obs.tau.ac.il/$\tilde{\;\:}$barkana/codes.html.} can
also be used to compute the final peculiar velocity profiles around
virialized halos as well. The total velocity of the gas is the sum
of the peculiar velocity and the Hubble expansion. In these
infalling regions around halos, we assume that the gas overdensity
and velocity field follow the dark matter perfectly, with the
scaling factor $\Omega_{\rm b}/\Omega_{\rm m}$.

\begin{figure}
\centering{\resizebox{8cm}{6cm}{\includegraphics{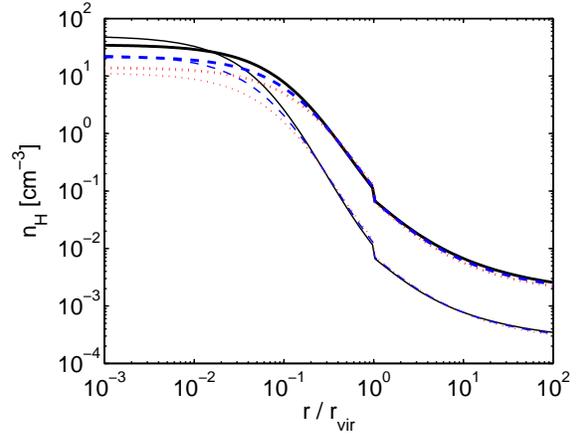}}}
\caption{The hydrogen number density profiles in and around dark
matter halos of different masses: $M = 10^6 M_\odot$ (solid curves),
$M = 10^7 M_\odot$ (dashed curves), and $M = 10^8 M_\odot$ (dotted
curves), respectively. The thick curves are for redshift 20, while
the thin curves are for redshift 10.} \label{Fig.nH}
\end{figure}

Several illustrative curves of the total hydrogen number density
profiles are shown in Fig.\ref{Fig.nH} for various halo masses and
two sets of redshifts. A constant hydrogen fraction of primordial
value $X_{\rm H} = 0.752$ is assumed here \citep{Spergel07}. The
halos with smaller mass and lower redshift are more concentrated. We
can see that there are discontinuities at $r_{\rm vir}$. That is
because when we adopt the gas density profile above, we are also
assuming that the virilization shock is located right at the virial
radius, and at the shock radius density and temperature jumps are
expected. The exact position of the shock radius is subject to
debate, and may depend on the assembly history of the halo. For an
cosmology with $\Omega_m = 1$, \citet{Bertschinger85} found that the
accretion shock forms at the radius $R_{90.8}$ which encloses a mean
dark matter density of 90.8 times the cosmic mean, while
\cite{Abel02} suggests that the shock radius is close to the virial
radius. Below we assume that the shock is located at the virial
radius, but note that the exact position of the shock radius has
only a small effect on the resulting 21 cm signal (see the next
section).

\subsection{Star Formation Criterion}\label{SBcriterion}
After the formation of dark matter halos, some of the halos could
form stars inside while others could not, depending mainly on the
cooling processes in each halo (see \citealt{MT08} for a recent
review). As for those halos with star formation, a star burst is
more likely than continuous formation at this early epoch, because,
small objects are very sensitive to feedbacks and therefore, as soon
as the first stars formed in the galaxy, their radiation and induced
supernovae could heat and eject the surrounding gas, quenching
subsequent star formation. In order to determine whether a halo is
able to undergo a star burst, here we use a timescale criterion for
star formation. If the time required for a halo to start forming
stars is longer than the Hubble time at the halo redshift, it will
remain starless, i.e. a system that is usually identified as a {\it
minihalo}. Otherwise, the halo has enough time to cool and collapse
to form stars, and becomes a {\it dwarf galaxy}. The timescale
required for turning on a star burst is modeled as the maximum
between the free-fall time $t_{\rm ff}$ and the cooling time $t_{\rm
cool}$, i.e. $t_{\rm SB} = \max \{\, t_{\rm ff},\, t_{\rm
cool}\,\}$, where $t_{\rm ff} = (3 \pi /\,32\, G \rho)^{1/2}$. As
for the minihalos and dwarf galaxies in the early universe, the main
coolant is the molecular hydrogen \citep{Abel02,Bromm02,Bromm04},
and $t_{\rm cool}$ is the H$_2$ cooling timescale. Defining $t_{\rm
cool} \equiv T/\dot T$, one can obtain \citep{Tegmark97}
\begin{equation}
t_{\rm cool} \,\approx\, 48,\!200\, \yr \left( 1 + \frac{10\,
T_3^{7/2}} {60 + T_3^4} \right)^{-1}\, e^{512\K/T} (f n_1)^{-1},
\end{equation}
where $T_3 \equiv T/10^3 \K$, $n_1 \equiv n_H / 1 \cm^{-3}$, and the
$f$ is the fraction of hydrogen in molecular form. This fraction
varies with time and to get the precise value at virialization it
would be necessary to follow the entire collapse history, but after
the virialization the H$_2$ fraction of a halo always saturates at
$\sim 10^{-3}$ \citep{Tegmark97}. For simplicity, we use $f =
10^{-3}$ hereafter.

The cooling and collapse processes start from the halo formation
time, which is usually defined as the time at which a halo first
acquires half of its final mass. There are several predictors for
the formation redshift distribution of dark matter halos, such as
the standard EPS model \citep{LC93}, the ellipsoidal collapse model
\citep{ST02}, and the non-spherical collapse boundary model
\citep{Chiueh01}. At high redshift, the EPS model fits the
simulation results best \citep{Lin03}. According to the EPS model
\citep{LC93}, the cumulative probability that a halo with mass $M_2$
at redshift $z_2$ has a formation time $t_{\rm F}$ prior to $t_1$ or
a formation redshift $z_{\rm F}$ higher than $z_1$ is
\begin{eqnarray}\label{Eq.cumuPzf}
P(t_{\rm F}<t_1|M_2,t_2) \,\equiv\, P(z_{\rm F}>z_1|M_2,z_2)
\nonumber \\
\,=\, \int_{S_2}^{S_h} \frac{M_2}{M_1}\,
f_{S_1}(S_1,\delta_{sc1}|S_2,\delta_{sc2})\, dS_1,
\end{eqnarray}
where $f_{S_1}(S_1,\delta_{sc1}|S_2,\delta_{sc2})\, dS_1$ is the
conditional probability function:
\begin{eqnarray}
f_{S_1}(S_1,\delta_{sc1}|S_2,\delta_{sc2})\, dS_1 \,=\,
\frac{\delta_{sc1} - \delta_{sc2}}{\sqrt{2\pi}\,(S_1-S_2)^{3/2}}
\nonumber \\
\,\times\, \exp{\left[-\,\frac{(\delta_{sc1} -
\delta_{sc2})^2}{2\,(S_1 - S_2)}\right]}\, dS_1 \; (S_1>S_2,
\delta_{sc1}>\delta_{sc2}).
\end{eqnarray}
Here $S$ is defined as the variance of density field smoothed on a
mass scale $M$, $S_1\equiv \sigma^2(M_1)$, $S_2\equiv
\sigma^2(M_2)$, $S_h\equiv \sigma^2(M_2/2)$, $\delta_{sc1} \equiv
\delta_{sc}(z_1)$, and $\delta_{sc2} \equiv \delta_{sc}(z_2)$. Given
any halo mass at the redshift of interest, we differentiate
Eq.(\ref{Eq.cumuPzf}) with respect to $z_{\rm F}$ to get the
differential distribution of the halo formation redshift, then its
formation redshift $z_{\rm F}$ is chosen by Monte-Carlo sampling of
the distribution curve.

The star burst time is simply $t_{\rm s} = t_{\rm F} + t_{\rm SB}$.
When the criterion $t_{\rm s} < t_{\rm H}$ is satisfied, a star
burst occurs at time $t_{\rm s}$, while $t_{\rm s} > t_{\rm H}$
implies that the halo is not able to form stars within a Hubble
time.

\subsection{The 21 cm Optical Depth}\label{OpticalDepthWithXRB}
With a high redshift GRB afterglow as the background source, the
neutral hydrogen in the IGM absorbs the 21 cm photons along the line
of sight. For the diffuse IGM beyond those regions affected by the
gravity of non-linear structures, it expands uniformly with the
Hubble flow, then the optical depth is \citep{Field59,Madau97,FOB06}
\begin{equation}
\tau_{\rm IGM} \,=\, \frac{3\,h_{\rm P}\,c^3 A_{10}}{32\, \pi\,
k_{\rm B}\,\nu_{10}^2}\, \frac{n_{\rm HI}(z)}{T_{\rm S}\,H(z)},
\end{equation}
where $h_{\rm P}$ is the Planck constant, $c$ is the speed of light,
$A_{10} = 2.85 \times 10^{-15} \psec$ is the Einstein coefficient
for the spontaneous decay of the 21 cm transition, and $\nu_{10} =
1420.4 \MHz$ is the rest frame frequency of the 21 cm transition.
Here $n_{\rm HI}$, $T_{\rm S}$, and $H$ are the neutral hydrogen
number density, spin temperature, and the Hubble parameter,
respectively.

In order to compute the $n_{\rm HI}$ and $T_{\rm S}$, a global
description of the ionization and thermal evolution of the IGM is
required. Several authors (e.g. \citealt{OH03,Furlanetto06}) have
shown that X-rays will likely heat the IGM significantly before
reionization completes. Also, if it existed, it could partially
ionize the IGM, suppress the formation of low mass minihalos, and
heat the infall regions around surviving minihalos and dwarf
galaxies. However, the heating processes are quite uncertain during
the early stages of reionization, and it is unknown what the X-ray
intensity should be at high redshifts. Here we include only the
X-ray background heating which is the most important heating
mechanism \citep{Furlanetto06}. Calibrating the X-ray ($>0.2\,
\keV$) luminosity of high-$z$ objects to nearby starburst galaxies,
we have \citep{Gilfanov04,Furlanetto06}
\begin{equation}
L_{\rm X} = 3.4 \times 10^{40}\, f_{\rm X}\, \left(\frac{\rm SFR}
{1\,M_{\rm \odot}\, \yr^{-1}} \right)\; \erg \psec,
\end{equation}
where $\rm SFR$ is the star formation rate, and $f_{\rm X}$ is a
correction factor accounting for the unknown properties of X-ray
emitting sources in the early universe. We will set $f_{\rm X}=0.1$
as the fiducial value as the early galaxies are probably much less
productive of X-rays than the nearby starburst galaxies, but we take
$f_{\rm X}$ as a free parameter and vary the level of X-ray
background by choosing different values of $f_{\rm X}$ to illustrate
the sensitivity of our results to this uncertain process. Assuming
that the star formation rate is proportional to the rate at which
matter collapses into galaxies \citep{Furlanetto06}, the total X-ray
emissivity $\epsilon_{\rm X}$ can be written as:
\begin{equation}
\frac{2}{3}\, \frac{\epsilon_{\rm X}}{k_{\rm B}\,n\,H(z)} =
5\times10^4 \K\, f_{\rm X}\, \left(\frac{f_\star}{0.1}\, \frac{{\rm
d} f_{\rm coll}/{\rm d}z}{0.01}\, \frac{1+z}{10}\right),
\end{equation}
where $n$ is the total particle number density (including hydrogen
atoms, protons, electrons, and helium), $f_{\rm coll}$ is the
fraction of matter in collapsed halos with $M > M_{\rm min}$, where
$M_{\rm min}$ is typically taken to be the mass threshold for atomic
cooling which has a virial temperature $T_{\rm vir} = 10^4 \K$, and
$f_\star$ is the star formation efficiency normalized to the cosmic
baryon content. We use a mass-dependent $f_{\star}$ provided by the
handy fit in \citet{SF09}, and evaluate it at $M_{\rm min}$ for the
most populous galaxies for approximation.

\begin{figure*}
\centering{\resizebox{17cm}{7cm}{\includegraphics{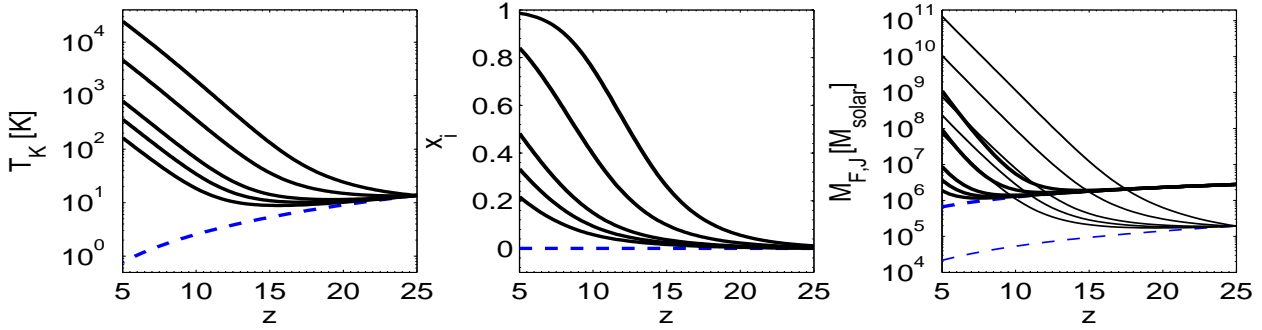}}}
\caption{The evolution of the global temperature ({\it left panel}),
the mean ionized fraction ({\it central panel}), and the filtering
mass and Jeans mass ({\it right panel}) of the IGM. The dashed
curves are for $f_{\rm X}=0$, and the solid curves from bottom to
top in each panel are for $f_{\rm X}=0.05$, $f_{\rm X}=0.1$, $f_{\rm
X}=0.2$, $f_{\rm X}=1.0$, and $f_{\rm X}=5.0$, respectively. In the
right panel, the thin and thick curves illustrate the evolutions of
the Jeans mass and the filtering mass respectively.}
\label{Fig.TIGM_xi_MFMJ}
\end{figure*}

A fraction of energy in the X-ray background contributes to the
heating of the IGM, and another fraction contributes to the
ionization of IGM, these fractions depend on the ionizing state of
the IGM \citep{VF08}. In the absence of an ionizing UV background,
the ionization would be caused mainly by these X-ray photons, and we
take into account both the primary and the secondary ionization
processes in computing this $\bar{x}_{\rm i}(z)$, assuming
ionization equilibrium. Then the evolution of the gas temperature in
the IGM can be described by \citep{Furlanetto06}
\begin{equation}\label{Eq.Tk}
\frac{{\rm d}T_{\rm K}}{{\rm d}t} \,=\, -\, 2\,H(z)\,T_{\rm K} +
\frac{2}{3}\, \frac{\epsilon_{\rm comp}}{k_{\rm B}n} +
\frac{2}{3}\, \frac{\epsilon_{\rm X,h}}{k_{\rm B}n},
\end{equation}
where the first item on the right-hand side describes the adiabatic
cooling due to Hubble expansion, $\epsilon_{\rm comp}$ is the
Compton heating/cooling rate per unit (physical) volume, and
$\epsilon_{\rm X,h}$ denotes the part of the total X-ray background
going into heating. The Compton term is (see e.g.
\citealt{Furlanetto06})
\begin{equation}
\frac{2}{3}\, \frac{\epsilon_{\rm comp}}{k_{\rm B}n} \,=\,
\frac{\bar{x}_{\rm i}}{1+f_{\rm He}+\bar{x}_{\rm i}}\,
\frac{8\,\sigma_{\rm T}\, u_\gamma}{3\, m_e\, c}\, (\,T_\gamma -
T_{\rm K}\,).
\end{equation}
Here $f_{\rm He}$ is the He fraction by number, $\sigma_{\rm T} =
6.65 \times 10^{-25} \cm^2$ is the Thomson scattering cross section,
$u_\gamma = (4\,\sigma/c)\,T_\gamma^4 \propto (1+z)^4$ is the energy
density of the CMB photons, where $\sigma = 5.67 \times 10^{-5}
\erg\, \cm^{-2} \psec \K^{-4}$ is the Stefan-Boltzmann constant, and
$m_e$ is the electron mass. The evolutions of the IGM temperature
and the mean ionized fraction caused by X-rays are shown in the left
and central panels of Fig.\ref{Fig.TIGM_xi_MFMJ} for several values
of $f_{\rm X}$. The case of $f_{\rm X}=0$ is denoted by the dashed
curves. It corresponds to the situation in which there is no X-ray
background, and the IGM temperature decreases adiabatically with a
mean ionized fraction of $3\times 10^{-4}$ which is the residual
electron fraction left over after the recombination. The solid
curves from bottom to top in each panel take $f_{\rm X}=0.05$,
$f_{\rm X}=0.1$, $f_{\rm X}=0.2$, $f_{\rm X}=1$, and $f_{\rm X}=5$,
respectively. Here we also illustrate the corresponding evolutions
of the Jeans mass (thin curves) and the filtering mass (thick
curves) in the right panel of Fig.\ref{Fig.TIGM_xi_MFMJ}. The
filtering mass, which provides a reasonable fit to the
characteristic mass, is $M_{\rm F}\sim 10^6 M_{\rm \odot}$ for
$f_{\rm X} \lesssim 0.2$, $M_{\rm F}\sim 2\times 10^6 M_{\rm \odot}$
for $f_{\rm X} \sim 1$, and $M_{\rm F}\sim 5\times 10^6 M_{\rm
\odot}$ for $f_{\rm X} \sim 5$ at $z\sim 10$.

The IGM creates a global decrement in the afterglow spectrum, on top
of which minihalos and dwarf galaxies produce deep and narrow
absorption lines. The main broadening mechanism of each absorption
line is the thermal broadening with the Doppler profile. The 21 cm
optical depth of an isolated object is the integral of the
absorption coefficient along the line of sight
\citep{Field59,Madau97,Furlanetto02}:
\begin{eqnarray}\label{Eq.tau}
\tau (\nu) &=& \frac{3\,h_{\rm P}\,c^3 A_{10}}{32\, \pi^{3/2}
k_{\rm B}}\, \frac{1}{\nu^2} \nonumber  \\
&\times& \int_{-\infty}^{+\infty} \frac{n_{\rm HI}(r)} {b(r)T_{\rm
S}(r)}\, \exp\left[\,-\, \frac{(u(\nu)-\bar
v(r))^2}{b^2(r)}\,\right] dx,
\end{eqnarray}
where $b(r)$ is the Doppler parameter of the gas, $b(r) =
\sqrt{\,2\,k_{\rm B}T_K(r)/m_{\rm H}}$, $u(\nu) \equiv c\,
(\nu-\nu_{10})/\nu_{10}$, and $\bar v(r)$ is bulk velocity of gas
projected to the line of sight at the radius $r$. Inside of the
virial radius, the gas is thermalized, and $\bar v(r) = 0$, while
the gas outside the virial radius has a bulk velocity contributed
from both the infall and the Hubble flow according to the ``Infall
Model''. The coordinate $x$ is related to the radius $r$ by $r^2 =
(\alpha\, r_{\rm vir})^2 + x^2$, where $\alpha$ is the impact
parameter of the penetrating line of sight in units of $r_{\rm
vir}$.

The spin temperature of neutral hydrogen is defined by the relative
occupation numbers of the two hyperfine structure levels, and it is
determined by three competing processes: (1) absorption of CMB
photons; (2) collisions with other hydrogen atoms, free electrons,
and other species; and (3) scattering with UV photons. The
equilibrium spin temperature is given by
\citep{Field58,FOB06}:
\begin{equation}
T_{\rm S}^{-1} \,=\, \frac{T_\gamma^{-1} + x_c\,T_{\rm K}^{-1} +
x_\alpha\, T_{\rm C}^{-1}} {1+x_c+x_\alpha},
\end{equation}
where $T_\gamma = 2.726(1+z)\,\K$ is the CMB temperature at redshift
$z$, $T_{\rm K}$ is the gas kinetic temperature, and $T_{\rm C}$ is
the effective color temperature of the UV radiation. In most cases,
$T_{\rm C}=T_{\rm K}$ due to the frequent Ly$\alpha$ scattering
\citep{FOB06}. The collisional coupling is described by the
coefficient $x_c$, and $x_\alpha$ is the coupling coefficient of the
Ly$\alpha$ pumping effect known as the Wouthuysen-Field coupling
\citep{Wouthuysen52,Field58}. The main contributions to $x_c$ are
H-H collisions and H-$e^-$ collisions, and it can be written as
\begin{equation}
x_c \,=\, x_c^{\rm eH} + x_c^{\rm H\!H} \,=\, \frac{n_e\,
\kappa_{10}^{\rm eH}}{A_{10}}\, \frac{T_\star}{T_\gamma} +
\frac{n_{\rm HI}\, \kappa_{10}^{\rm H\!H}}{A_{10}}\,
\frac{T_\star}{T_\gamma},
\end{equation}
where $T_\star = 0.0682\, \K$ is the equivalent temperature of the
energy slitting of the 21 cm transition, and $\kappa_{10}^{\rm eH}$
and $\kappa_{10}^{\rm H\!H}$ are the de-excitation rate coefficients
in collisions with free electrons and hydrogen atoms respectively.
These two coefficients at different temperatures are tabulated in
\citet{FOB06}. The coupling coefficient $x_\alpha$ is proportional
to the total scattering rate between Ly$\alpha$ photons and hydrogen
atoms,
\begin{equation}
x_\alpha \,=\, \frac{4\,P_\alpha}{27\,A_{10}}\,
\frac{T_\star}{T_\gamma},
\end{equation}
where the scattering rate $P_\alpha$ is given by
\begin{equation}
P_\alpha \,=\, c\,\sigma_\alpha\, \frac{n_\alpha^{\rm
tot}}{\Delta\nu_D} \,=\, 4\pi\, \sigma_\alpha J_\alpha.
\end{equation}
Here $\sigma_\alpha \equiv {\displaystyle \frac{\pi e^2}{m_e c}
f_\alpha}$ where $f_\alpha=0.4162$ is the oscillator strength of the
Ly$\alpha$ transition, $n_\alpha^{\rm tot}$ is the total number
density of Ly$\alpha$ photons, $J_\alpha$ is the number intensity of
the Ly$\alpha$ photons, and $\Delta\nu_D = (b/c)\,\nu_\alpha$ is the
Doppler width with $b$ being the Doppler parameter and $\nu_\alpha$
being the Ly$\alpha$ frequency.

In addition to the global $\bar{x}_{\rm i}(z)$ and $T_{\rm IGM}(z)$,
to compute the line profiles of the 21 cm absorptions by minihalos
and dwarf galaxies, we need a detailed prescription of the
ionization state, the temperature profile, and the Ly$\alpha$ photon
density in and around these objects. We model respectively these
properties of both minihalos and dwarf galaxies, as well as the
intensity of Ly$\alpha$ background in the following.

\subsubsection{Modeling the minihalos}
Minihalos refer to those small halos that are not capable of
hosting stars, and they represent a very numerous population
according to the halo mass function. As we are considering the
early stages of reionization, when the IGM was slightly ionized
with some rare sites illuminated by the earliest galaxies, the
background radiation field has not been set up. However, the gas
in minihalos, which has much higher density, could be
collisionally ionized partially, depending on its temperature. The
HI fraction is computed from collisional ionization equilibrium
(CIE)
\begin{equation}
n_e\,n_{\rm HI}\, \gamma \:=\: \alpha_{\rm B}\,n_e\,n_p,
\end{equation}
where $\gamma$ is the collisional ionization coefficient
\citep{Cen92}, and $\alpha_{\rm B}$ is the case B recombination
coefficient \citep{HG97} which is appropriate for the high-z and
low-mass halos as well as the IGM that are far from fully ionized.

\begin{figure}
\centering{\resizebox{8cm}{6cm}{\includegraphics{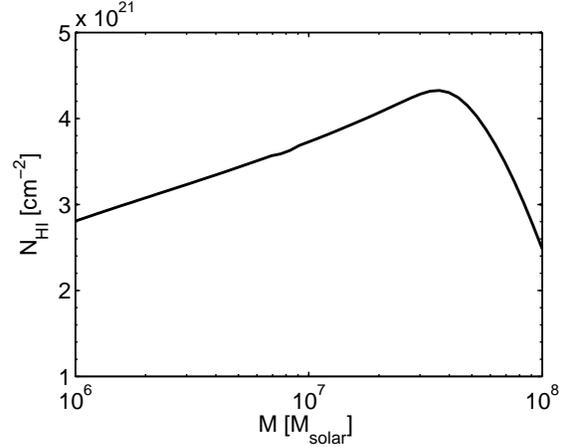}}}
\caption{The HI column density as a function of minihalo mass at
redshift 10.} \label{Fig.N_HI}
\end{figure}

The coefficients $\gamma$ and $\alpha_{\rm B}$ are both functions
of temperature. During the virialization process, the gas inside a
minihalo is shock heated to the virial temperature. As for the gas
outside the halo ($r>r_{\rm vir}$), since we are looking at high
redshifts when reionization is still very patchy, the IGM around a
minihalo might be much colder or much hotter than the gas inside
the minihalo depending on its environment. In other words, either
the minihalo is embedded in an almost neutral region, or it is
embedded in an HII region produced by a neighboring galaxy. If we
go to high enough redshifts, the probability for the halo to be in
a neutral region increases. With an X-ray background, we assume
that the temperature of gas around each minihalo equals to $T_{\rm
IGM}$.
The HI column density of a minihalo (integrated from $-r_{\rm vir}$
to $+r_{\rm vir}$ along a line of sight passing through the halo
center) as a function of its mass is shown in Fig.\ref{Fig.N_HI} for
redshift 10. For low mass minihalos, the HI column density increases
with the halo mass, but it decreases rapidly for halo masses larger
than $\sim 5\times 10^7 M_{\odot}$, as the gas is gradually ionized
in those halos with higher virial temperatures.

Another ingredient for the computation of 21 cm optical depth is the
number density of Ly$\alpha$ photons, which is essential for
determining the spin temperature of hydrogen. The background
Ly$\alpha$ photons can not penetrate into minihalos and contributing to the
Ly$\alpha$ flux inside them (see the arguments in section
\ref{LyaBackground}), and in the absence of stars, the Ly$\alpha$
photons could come only from recombinations in the gas itself. The
Ly$\alpha$ production rate is
\begin{equation}
\dot n_{\rm Ly\alpha} \,=\, (2/3)\,\alpha_{\rm B}\,n_e\, n_p.
\end{equation}
The factor of two-thirds accounts for the probability of the
recombination leading to a Ly$\alpha$ photon and the other
one-third is the probability of obtaining photons of frequencies
different from the Ly$\alpha$ \citep{Osterbrock89}. Because of the
extremely large cross-section of Ly$\alpha$ resonant scattering,
the Ly$\alpha$ photons cannot escape immediately from the halo,
but scatter with the hydrogen atoms frequently and diffuse in the
gas. This process is well described as a ``random walk''. Thus,
the Ly$\alpha$ photons can accumulate in a given region for a
diffusion time, $t_{\rm diffu} = l_{\rm rms}^2/(c\,l_{\rm mfp})$,
where $l_{\rm rms}$ is the r.m.s. distance that Ly$\alpha$ photons
have to travel in order to escape from the region, and $l_{\rm
mfp} = (n_{\rm HI}\, \sigma_\alpha/\Delta\nu_D)^{-1}$ is the mean
free path of the Ly$\alpha$ photons. For the case of minihalo
$l_{\rm rms}=r_{\rm vir}$. However, the Ly$\alpha$ accumulating
time $t_{\rm acc}$ is also limited by the time interval between
the halo formation and the Hubble time, i.e. $(t_{\rm H} - t_{\rm
F})$, then the real accumulating time of Ly$\alpha$ photons is
$t_{\rm acc}=\min\{\,t_{\rm diffu},\, t_{\rm H}-t_{\rm F}\,\}$.
Actually, $t_{\rm diffu}$ is quite large, and it is always longer
than the Hubble time, so the Ly$\alpha$ photons would accumulate
for a time interval $(t_{\rm H} - t_{\rm F})$. Due to the large
accumulating time, even with a low ionized fraction in the IGM,
the Ly$\alpha$ photons from recombination could also be important
outside the $r_{\rm vir}$ as compared to the background Ly$\alpha$
photons.

\begin{figure}
\centering{\resizebox{8cm}{10cm}{\includegraphics{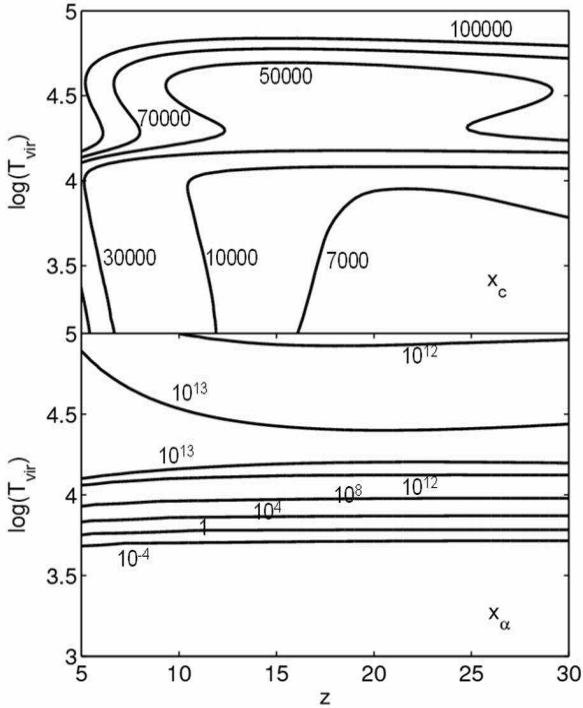}}}
\caption{Contours of collisional coupling coefficient $x_c$ (upper
panel) and Ly$\alpha$ coupling coefficient $x_\alpha$ (bottom panel)
on $\log (T_{\rm vir}) - z$ planes. From bottom to top in each
panel, the values of the contours are $7\times10^3$, $10^4$,
$3\times10^4$, $5\times10^4$, $7\times10^4$, and $10^5$ for $x_c$,
and $10^{-4}$, $1$, $10^4$, $10^8$, $10^{12}$, $10^{13}$, $10^{13}$,
and $10^{12}$ for $x_\alpha$, respectively.} \label{Fig.xaxc}
\end{figure}

In order to illustrate the relative effects of collisional coupling
and Ly$\alpha$ coupling in minihalos, we plot several contours of
the coefficients $x_c$ (upper panel) and $x_\alpha$ (bottom panel) on
$\log (T_{\rm vir}) - z$ planes in Fig.\ref{Fig.xaxc}. These
coefficients are all calculated for the central properties of halos,
and to show clearly the dependence of $x_\alpha$ on the virial
temperature of halos, we set the accumulation time of the Ly$\alpha$
photons to be a Hubble time at each redshift.

As we see clearly from the contours,
$x_\alpha$ dominates over $x_c$ for high mass halos, in which gas is
collisionally ionized and Ly$\alpha$ photons are effectively
produced and accumulated from recombinations. On the other hand,
$x_c$ dominates over $x_\alpha$ for low mass halos which are almost
completely neutral.

As for $x_\alpha$, it has a very weak dependence on the redshift,
but increases dramatically around $T_{\rm vir}\sim10^4\K$ due to the
change of ionized fraction there. At higher temperatures, when the
gas is almost collisionally ionized, the $x_\alpha$ slowly drops
again because of the decreasing $\alpha_{\rm B}$ with temperature.

There are several effects taking place to determine the behavior of
$x_c$. In general, $x_c$ increases with temperature because of the
increasing collisional de-excitation rate $\kappa_{10}^{\rm H\!H}$
and $\kappa_{10}^{\rm eH}$. At low temperatures, $x_c$ comes mainly
from the H-H collisions, and the slowly decreasing $x_c$ with
increasing redshift is caused by the decreasing central density of
hydrogen in less concentrated minihalos at higher redshifts. The
coupling changes from H-H collisions-dominated to $e^-$-H
collisions-dominated as the temperature goes up, and this is in part
responsible for the non-monotonic behavior of $x_c$ at $\log (T_{\rm
vir}) \sim 4-4.5$. In addition, given a redshift, the concentration
is lower for higher mass (higher $T_{\rm vir}$) halos, and the
central density of gas is lower. Therefore, $n_e$ is smaller at high
temperatures when the gas is almost ionized and $n_e$ represents the
total hydrogen density. This could also result in the non-monotonic
behavior of $x_c$.

\subsubsection{Modeling the dwarf galaxies}
As discussed in section \ref{SBcriterion}, star bursts could occur
in some of the halos, turning them into dwarf galaxies when the
cooling process allows the gas to cool and collapse within a Hubble
time. Depending on the initial mass function (IMF) and star
formation efficiency, the stars  produce a radiation field
that will photonionize the gas creating an HII region, heat the gas
around, and produce Ly$\alpha$ photons at the same time. All these
effects could have influences on the strength and shape of the 21 cm
signal of the galaxy.

The initial mass function and the emission spectrum of the first
galaxies remain very uncertain (see \citealt{CFreview05} for a
review). However, \citet{Schaerer02,Schaerer03} has examined
spectral properties of the ionizing continua of high redshift
starburst galaxies for various IMFs, metallicities and star
formation histories\footnote{Data are available at:\\
http://cdsarc.u-strasbg.fr/cgi-bin/Cat?VI/109}. We make use of
their results for the emitting rate of H and He$^+$ ionizing
photons per solar mass of burst stars, denoted by $Q_{\rm H}$ and
$Q_{\rm He^+}$, as well as the average energy per photon with
$E>54.4\eV$ (He$^+$ ionizing threshold) denoted by $\bar E_{\rm
He^+}$. The Lyman continuum photons are divided into two parts:
photons with $E<100 \eV$ could effectively result in
photonionization, and those soft X-rays with $E>100 \eV$ have a
large probability to escape, and serve as an extra heating
mechanism in addition to the X-ray background outside the HII
region. Assuming a simple power law for the high energy tail of
the emission spectrum, we extrapolate $Q_{\rm He^+}$ and $\bar
E_{\rm He^+}$ to get the production rate of soft X-rays $Q_{\rm
X}$ and their average photon energy $\bar E_{\rm X}$. Thus the
production rate of ionizing photons is $Q_{\rm ion} = Q_{\rm H} -
Q_{\rm X}$.

As we are considering an early reionization epoch when the UV
background has not been set up, all the ionizing photons creating an
HII region come from the stellar sources inside of the dwarf galaxy,
i.e. the $Q_{\rm ion}$. Considering the soft spectra of stellar
sources, we assume a sharp boundary for each HII region, i.e. $x_i =
1$ for $r \le R_{\rm HII}$, 
$x_i = \bar{x}_i(z)$ for $r>R_{\rm HII}$ and $r>r_{\rm vir}$, and
$x_i = x_i({\rm CIE})$ for $R_{\rm HII}<r\le r_{\rm vir}$ when
$R_{\rm HII}< r_{\rm vir}$. Further, we assume all the stars to be
located at the center of the galaxy, and neglect the Hubble
expansion during the growth of the HII region, as its growth time
is much shorter than the Hubble time. Starting from the star burst
time, $t_*$, the evolution equation of the radius of an HII region
is
\begin{equation}\label{Eq.RHII}
n_{\rm HI}(R)\,4\pi R^2\,\frac{dR}{dt} \,=\, \dot N_{\gamma}(t) -
\int_0^{R(t)} \alpha_{\rm B}\, n_e\, n_p\, 4\pi r^2 d{\rm r},
\end{equation}
where $\dot N_{\gamma}(t)$ is the ionizing rate, and the second term
on the right-hand side accounts for recombinations in the HII
region. Here $\alpha_{\rm B} = \alpha_{\rm B}(10^4 \K) = 2.59 \times
10^{-13} \cm^3 \psec$ is used, as appropriate for the temperature of
an ionized region. Here we negelect the change in gas density profile due
to the heating, as we will see later that it is quite weak.
Before the ionizing front breaks out of the virial radius, the ionization is
caused by ionizing photons that did not escape the galaxy, then
\begin{equation}
\dot N_{\gamma}(t) \,=\, (1-f_{\rm esc})\, Q_{\rm ion}\, f_{\star}\,
\frac{\Omega_b}{\Omega_m}\, M,
\end{equation}
where $f_{\rm esc}$ is the escape fraction of ionizing photons, and
$f_{\star}$ is the star formation efficiency normalized to the
cosmic baryon content associated with a halo of mass $M$. We use
$f_{\rm esc}=0.07$ as favored by the early reionization model (ERM,
\citealt{Gallerani08}), and a mass-dependent $f_{\star}$ provided by
the handy fit in \citet{SF09}. After the ionizing front grows larger
than the virial radius, all the ionizing photons have escaped, then
$f_{\rm esc} = 1$, and
\begin{equation}
\dot N_{\gamma}(t) \,=\, f_{\rm esc}\, Q_{\rm ion}\, f_{\star}\,
\frac{\Omega_b}{\Omega_m}\, M.
\end{equation}

\begin{figure}
\centering{\resizebox{8cm}{6cm}{\includegraphics{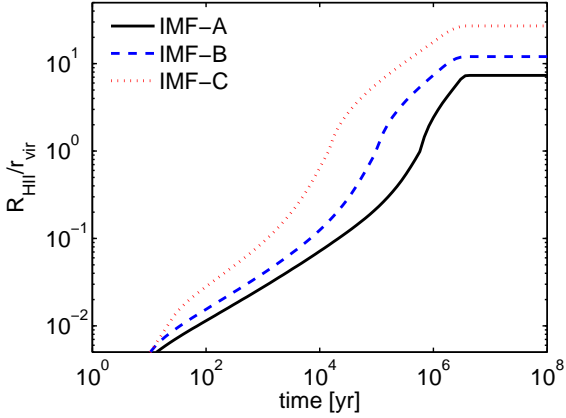}}}
\caption{The growth of $R_{\rm HII}$ for a dwarf galaxy with mass $M
= 10^7 M_\odot$ and redshift $z = 10$ after a star burst. The solid
(black), dashed (blue), and dotted (red) curves are for IMF-A (1 --
100 $M_\odot$), IMF-B (1 -- 500 $M_\odot$), and IMF-C (50 -- 500
$M_\odot$), respectively. All the IMF models have a Salpeter shape,
the metallicity shown here is $Z = 10^{-7}$, and we set $f_{\rm
X}=0.1$ for the X-ray background.} \label{Fig.RHIIevol}
\end{figure}

We plot the radius of the HII region as a function of the time after
star burst for a dwarf galaxy with $M=10^7 M_\odot$ at redshift 10
in Fig.\ref{Fig.RHIIevol}. The (physical) virial radius of this
galaxy is $r_{\rm vir} = 641 \pc$. 
Since $Q_{\rm H}$ has a weak dependence on the metallicity but very
sensitive to the IMF, we fix the metallicity at Z = $10^{-7}$, and
show the growth of $R_{\rm HII}$ for three different IMFs. These IMF
models are correspondingly chosen after
\citet{Schaerer02,Schaerer03}, with a power-law slope of the
Salpeter value ($\alpha$ = 2.35) and different upper and lower mass
limits: 1 -- 100 $M_\odot$ (IMF-A), 1 -- 500 $M_\odot$ (IMF-B), and
50 -- 500 $M_\odot$ (IMF-C). The $R_{\rm HII}$ grows very fast at
the beginning, but limited by the speed of light. At later stages,
it is also limited by the stellar lifetime. As the stars fade away,
the subsequent supernova explosions will sweep the surrounding gas
away, thus suppressing the 21 cm absorption signal. For this reason
we stop the evolution of the HII radius at its maximum value. As one
could expect, the size of an HII region is very sensitive to the IMF
due to the strong dependence of Lyman continuum production on
stellar mass. Here we are considering isolated HII regions around
dwarf galaxies, but in reality the galaxies will be somewhat
clustered. So the ionized regions may extend further. However, the
clustering of the first galaxies is beyond the scope of this work,
and we reserve this to future works.

Once an HII region is created, the gas inside rapidly approaches
the temperature around $2\times10^4\K$ \citep{Meiksin07}. For the
21 cm absorption which requires neutral hydrogen, it is more
important to compute the gas temperature outside the HII region.
Here we include the heating by local soft X-rays emitted by the
dwarf galaxies. 
While the stars in the galaxy are also emitting Ly$\alpha$
photons, the heating effect due to the repeated Ly$\alpha$
scattering is negligible \citep{ChenM04}.
So the evolution of gas temperature can still be described by
Eq.(\ref{Eq.Tk}), but with an extra contribution to the
$\epsilon_{\rm X,h}$ from the local X-rays. The energy deposition
rate of local soft X-ray heating is
\begin{equation}
\epsilon_{\rm X,g} \,=\, \mathcal{R}\, n_{\rm HI}\, (\bar E_{\rm
X} - E_{\rm th})\, f_{\rm X,h},
\end{equation}
where $\mathcal{R}$ is the ionizing rate by the soft X-rays, $E_{\rm
th}=13.6\eV$ is the ionization threshold of hydrogen, and $f_{\rm
X,h}$ is the fraction of the primary electron energy deposited into
heat. The X-ray ionizing rate can be written as $\mathcal{R} =
Q_{\rm X}(t)\,M_\star\, \sigma_{\rm I}(\nu)/(4\pi r^2)$, where
$M_\star = f_\star\,(\Omega_b/\Omega_m)\,M$, and $\sigma_{\rm
I}(\nu)$ is the photonionization cross section of hydrogen
\citep{Meiksin07}:
\begin{equation}
\sigma_{\rm I}(\nu) \,=\, \sigma_0 \left[\beta
\left(\frac{\nu}{\nu_{\rm th}}\right)^{-s} +
(1-\beta)\left(\frac{\nu}{\nu_{\rm th}}\right)^{-s-1}\right],
\end{equation}
where $\sigma_0 = 6.30 \times 10^{-18} \cm^2$, $\nu_{\rm th} = 3.290
\times 10^{15} \Hz$ corresponding to the ionization threshold
13.6eV, $\beta = 1.34$, and $s = 2.99$. The average frequency of
photons with $E \ge 0.1\keV$ is used here. As for $f_{\rm X,h}$, we
use a handy fit by \citet{VF08}, which is a function of ionized
fraction $x_i$.

\begin{figure}
\centering{\resizebox{8cm}{6cm}{\includegraphics{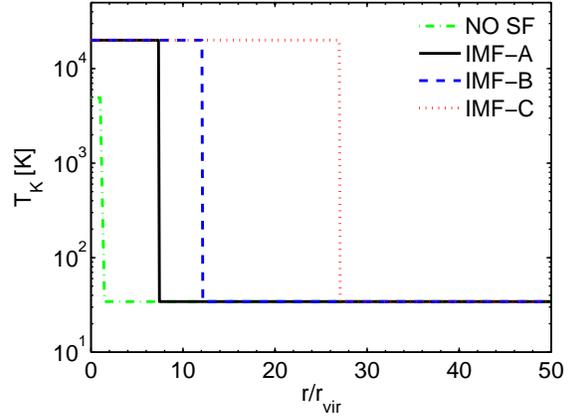}}}
\caption{The temperature profiles of gas around a dwarf galaxy with
mass $M = 10^7$ and metallicity $Z = 10^{-7}$ at redshift $z = 10$.
The solid (black), dashed (blue), and dotted (red) curves are for
Salpeter IMFs with mass ranges of 1 -- 100 $M_\odot$ (IMF-A), 1 --
500 $M_\odot$ (IMF-B), and 50 -- 500 $M_\odot$ (IMF-C),
respectively. The temperature profile of gas if there is no star
formation is also plotted with the dot-dashed (green) curve, and a
fiducial value of $f_{\rm X} = 0.1$ is assumed for the X-ray
background.} \label{Fig.Tk}
\end{figure}

With the initial condition of the IGM temperature at the star burst
time, we solve for the gas temperature around an HII region, and
plot the results for a dwarf galaxy with $M=10^7 M_\odot$ at
redshift 10 in Fig.\ref{Fig.Tk}. We take $f_{\rm X} = 0.1$ as the
fiducial value for the X-ray background. The solid, dashed, and
dotted curves are for the three IMF models used in
Fig.\ref{Fig.RHIIevol}. For illustration, we have assumed that the
star burst occurred at the same time as the halo formation, and the
temperatures are all evaluated at $t_{\rm value} = t_{\rm H}-t_{\rm
F}$, where the most probable value for $z_{\rm F}$ is used. For
comparison, the temperature profile for a starless minihalo of the
same mass and redshift is also shown as the dot-dashed line. Inside
the HII region, the temperature is fixed at $2 \times 10^4$ K, and
it drops down very sharply at the ionizing front due to the lower
heating rate outside, as we could expect for stellar sources. The
main heating mechanism outside $r_{\rm vir}$ is the X-ray background
heating.

Apart from ionizing photons and soft X-rays, the dwarf galaxy is
also producing Ly$\alpha$ photons which are essential for
determining the spin temperature of hydrogen. However, those
Ly$\alpha$ photons directly emitted by the galaxy could hardly
contribute to the coupling in the gas outside the HII region due to
the extremely long diffusion time
, and they will be blocked near the ionization front. It is the
Ly$\alpha$ photons cascaded from soft X-rays that penetrate into the
nearby IGM and effectively couple the spin temperature to the
kinetic temperature of the gas \citep{ChenM08}. So the contributing
Ly$\alpha$ photons near an HII region come from recombination, the
Ly$\alpha$ background, as well as the soft X-ray cascading.
\citet{VF08} have examined the fraction $f_{\rm Ly\alpha}$ of the
primary energy of electrons ionized by X-rays that converts into
Ly$\alpha$ radiation. Using their result, the Ly$\alpha$ production
rate from cascading around the HII region is
\begin{equation}
\dot n_{\rm Ly\alpha} \,=\, \mathcal{R}\, n_{\rm HI}\,\frac{\bar
E_{\rm X} - E_{\rm th}}{h\nu_{\alpha}}\, f_{\rm Ly\alpha},
\end{equation}
where $\nu_{\alpha} = 2.47\times10^{15} \Hz$ is the frequency of
Ly$\alpha$ photons. The total number density of Ly$\alpha$ photons
is obtained by integrating $\dot n_{\rm Ly\alpha}$ over an
accumulating time $t_{\rm acc}$.

In this case, the accumulating time of Ly$\alpha$ photons is $t_{\rm
acc} = \min\{\,t_{\rm diffu}, t_{\rm Xray}, (t_{\rm H}-t_{\rm
s})\,\}$, where $t_{\rm diffu}$ is the Ly$\alpha$ diffusion time as
before, but we take the mean free path (m.f.p.) of the soft X-rays
as the $l_{\rm rms}$ here. $t_{\rm Xray}$ is the time during which
the dwarf galaxy is producing X-rays, and $(t_{\rm H}-t_{\rm s})$ is
the time interval between the halo redshift and the star burst
redshift. Because of the large m.f.p. of the X-rays (compared to the
$r_{\rm vir}$), the Ly$\alpha$ diffusion time is also very large in
this case, even larger than the Hubble time. As a result, the
Ly$\alpha$ accumulating time is determined by $(t_{\rm H}-t_{\rm
s})$ for the cases with IMF-A and IMF-B, while for the case with
IMF-C, the Ly$\alpha$ accumulating time is limited by the lifetime
of the stars.

\subsubsection{The Ly$\alpha$ background}\label{LyaBackground}
In addition to the Ly$\alpha$ photons produced by recombination
and those cascaded Ly$\alpha$ photons near galaxies, there is also
a Ly$\alpha$ background. The Ly$\alpha$ background originates from
the continuum photons to the blue side of the Ly$\alpha$ line
emitted by the early galaxies, and its intensity becomes important
at very high redshift \citep{ChenM08}. Only photons with $\nu <
\nu_\beta$ can be redshifted to $\nu_\alpha$, and all photons of
higher frequencies will be absorbed at the Ly$\beta$ or higher
resonance Lyman series lines. Let $E(\nu)$ be the number of
photons emitted over stars' lifetime per baryon in one of the
halos that collapse to form stars and per unit of frequency.
According to the stellar spectra we adopted from
\citet{Schaerer03}, $E(\nu)$ is well approximated as flat near the
Ly$\alpha$ frequency, then the intensity of the continuum
Ly$\alpha$ background can be written as \citep{ChenM08}
\begin{equation}\label{Eq.J_alpha}
J_\alpha \,=\, \tilde{J}_0 \,\bar{E}_\alpha \nu_\alpha \,
[F(z)-F(z_\beta)].
\end{equation}
Here $\tilde{J}_0$ is the fiducial Ly$\alpha$ intensity
corresponding to a photon density of one photon per hydrogen atom
per log frequency,
\begin{equation}
\tilde{J}_0 \,=\, \frac{c\,n_{\rm H}}{4\pi\,\nu_\alpha},
\end{equation}
where $n_{\rm H}$ is the number density of hydrogen. $\bar{E}_\alpha
= \bar{E}(\nu_\alpha)$ is averaged over all the galaxies in the mass
range of our consideration, and we estimate this value from the
spectrum data provided by \citet{Schaerer03}. In
Eq.(\ref{Eq.J_alpha}), $F(z)$ is the fraction of mass bound in
star-forming halos,
\begin{equation}
F(z)\,=\, \int_{M_{\rm min}}^{M_{\rm max}} \frac{M}{\bar\rho_0}\,
n(M,z)\, p_\star(M,z)\, dM,
\end{equation}
where $n(M,z)$ is the halo mass function and $p_\star(M,z)$ is the
probability of having star formation for a halo with mass $M$ at
redshift $z$. We derive $p_\star(M,z)$ according to the star
formation criterion given in Section \ref{SBcriterion}. At $z=20$,
$J_\alpha \approx 3 \times 10^{-11}\, \cm^{-2} \psec \psr \Hz^{-1}$,
while at $z=10$, it has increased to $J_\alpha \approx 2 \times
10^{-10}\, \cm^{-2} \psec \psr \Hz^{-1}$ which is already important
for the Wouthuysen-Field coupling effect. 

The problem of Ly$\alpha$ background propagating into minihalos is
complicated. The situation in minihalos would be different from that
in the IGM because of their higher column densities of neutral
hydrogen (typically, $N_{\rm HI} \sim 10^{21}\, \cm^{-2}$). The
Ly$\alpha$ scattering cross section in a region of high density
consists of a Doppler core with a typical width of $\Delta\nu_{\rm
D} = b/\lambda_\alpha$ centered on the Ly$\alpha$ frequency, and
Lorentzian wings outside the core. Once a photon is redshifted or
scattered into the ``core'' frequencies, the cross section is large
and its mean free path becomes very short and is spatially confined
within a small region with very long diffusion time. Thus, the
photons which eventually make up the Ly$\alpha$ background come from
the blue side of the Ly$\alpha$ line, at the edge of the ``core''
and ``wing''. In the case of minihalo, however, these photons would
be blocked at the surface of the minihalo. The redshift across the
minihalo is also too small for bluer wing photons which is not
blocked at the edge of the minihalo to get redshifted into the
core-wing boundary region when they arrive at the center of the
halo.

Another possibility is to consider photons with such frequency that
their optical depth through the minihalo is about 1. Such a photon
would scatter once inside the halo. One scattering per photon is of
course negligible for the coupling effect in itself, but the
frequency of the photon would be changed after the scattering. If
the frequency increases, then the photon cross section become even
smaller and would escape right away, but if the photon loses energy
during the collision and shifts to lower frequency, the cross
section would be increased. Could such a change bring the photon
from the ``wing'' to the ``core''? The frequency $\nu$ of such a
wing photon is given by $N_{\rm HI}(r_{\rm vir})\,\sigma_\alpha(\nu)
= 1$, where $N_{\rm HI}(r_{\rm vir})$ is the column density of the
minihalo from $r_{\rm vir}$ to its center, and $\sigma_\alpha(\nu)$
is the scattering cross section on the Ly$\alpha$ damping wing. For
typical thermal velocity of the H atom in the minihalo, we estimate
that the change of photon frequency in one scattering ${\rm d}\,\nu
<2.2\times10^{11} \Hz$, comparable to the size of the core. On the
other hand, the frequency distance to the line center for such a
photon is $(\nu-\nu_\alpha) \sim 10^{13} \Hz$, so ${\rm d}\,\nu \sim
1\%(\nu-\nu_\alpha)$. Therefore, during one or even a few favorable
scatterings the photon could not enter the core and should escape
away. We conclude that with the exception of the surface, the
Ly$\alpha$ background could not affect the spin temperatures of the
minihalos.

\section{THE 21 CM LINE PROFILES}\label{profiles}
With the detailed models of gas density, velocity profile,
ionization state, temperature evolution, and the intensity of both
local and global Ly$\alpha$ photons for the minihalos and the
dwarf galaxies described above, we are now ready to compute their
spin temperature profiles and 21 cm optical depths. In this
section, we show our results of these detailed profiles for a
variety of parameters. Although it will be very challenging to
resolve them with radio instruments in the near future, these
analyses help us explore the physical origins of the line profiles
and understand the physics behind.

\subsection{The Coupling Effects and Spin Temperature}

\begin{figure*}
\begin{tabular}{cc}
\includegraphics[totalheight=6cm,width=7.5cm]{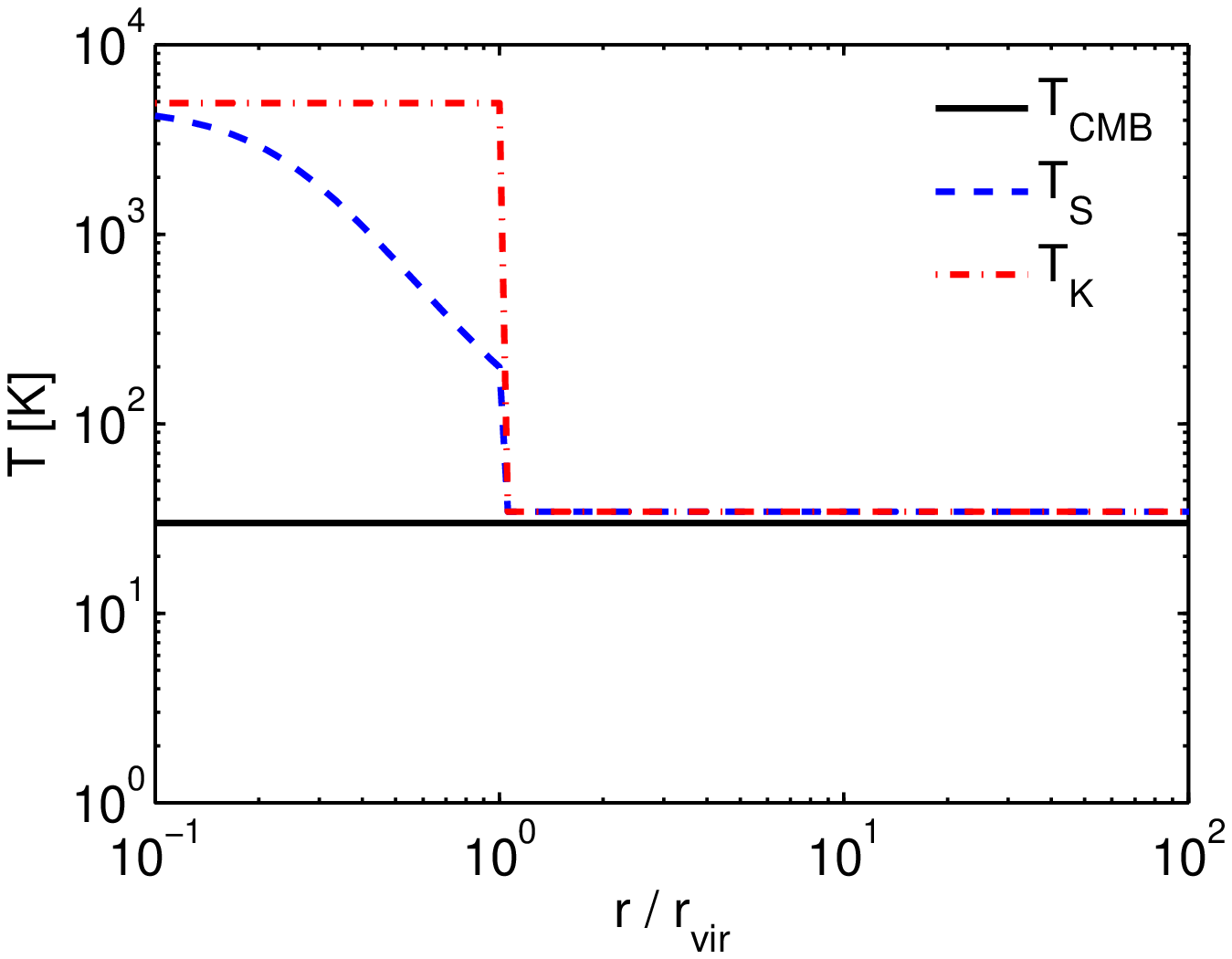}
&\includegraphics[totalheight=6cm,width=7.5cm]{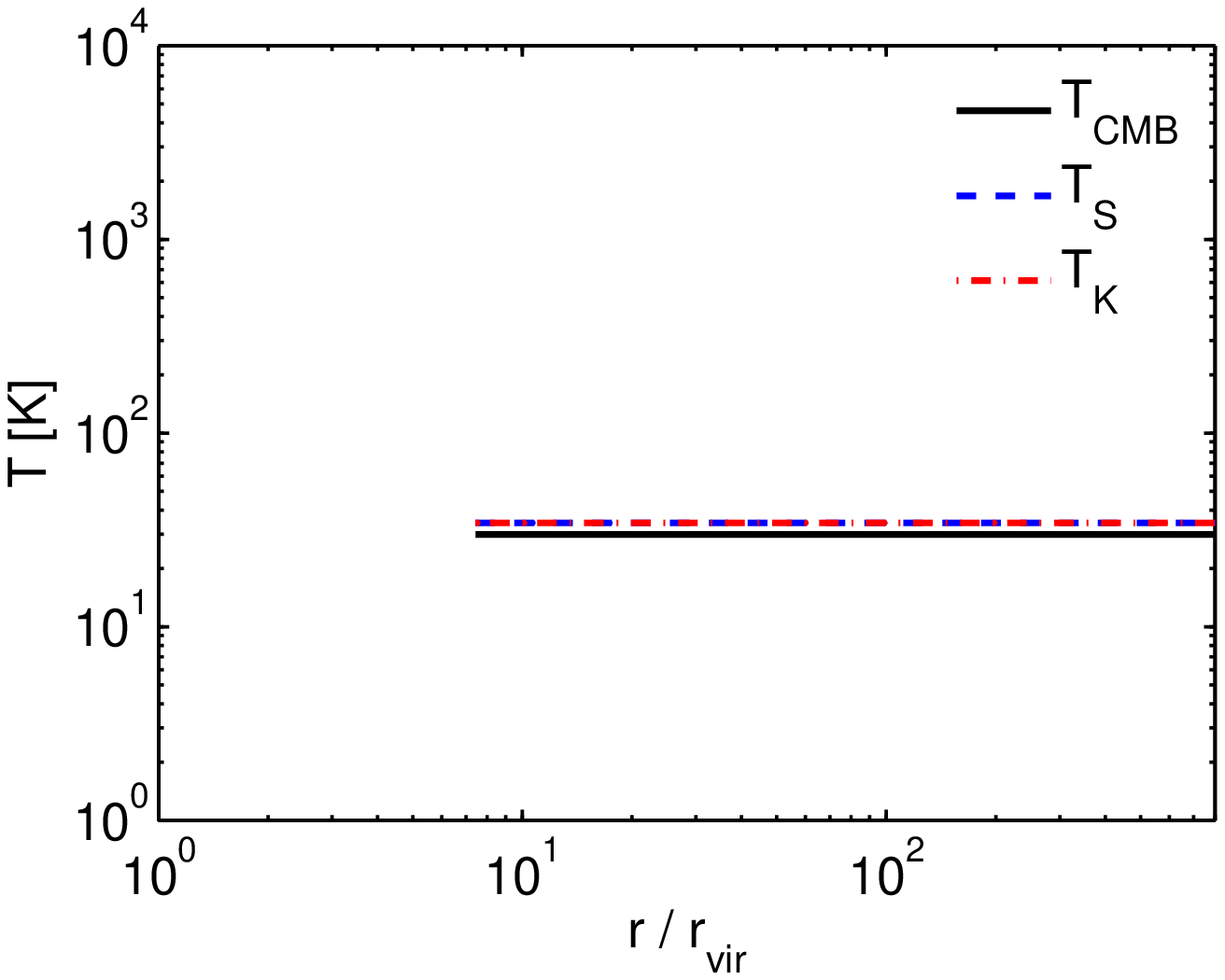}
\end{tabular}
\caption{Spin temperature (dashed lines) and kinetic temperature
(dot-dashed lines) profiles of a minihalo/dwarf galaxy with mass
$M = 10^7 M_\odot$ at redshift 10. A fiducial value of $f_{\rm
X}=0.1$ is assumed for the X-ray background. The solid curves
represent the CMB temperature. {\it Left:} the case for a minihalo
in CIE. {\it Right:} the case for a dwarf galaxy which is
photonionized after a star burst with IMF model-A and metallicity
$Z = 10^{-7}$. The temperature curves are cut at the HII radius of
7.02 $r_{\rm vir}$. } \label{Fig.TsTk}
\end{figure*}

\begin{figure*}
\begin{tabular}{cc}
\includegraphics[totalheight=6cm,width=7.5cm]{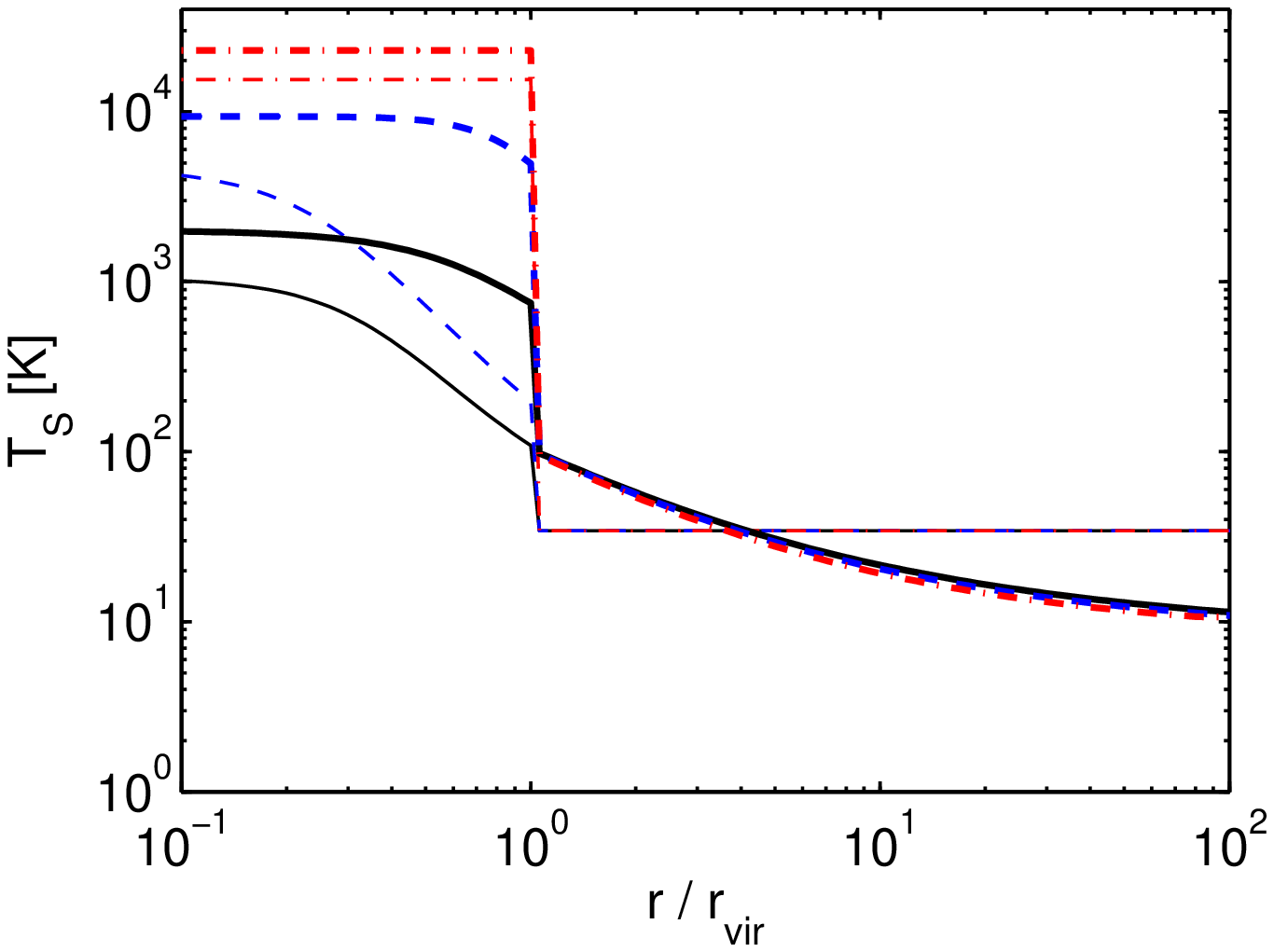}
&\includegraphics[totalheight=6cm,width=7.5cm]{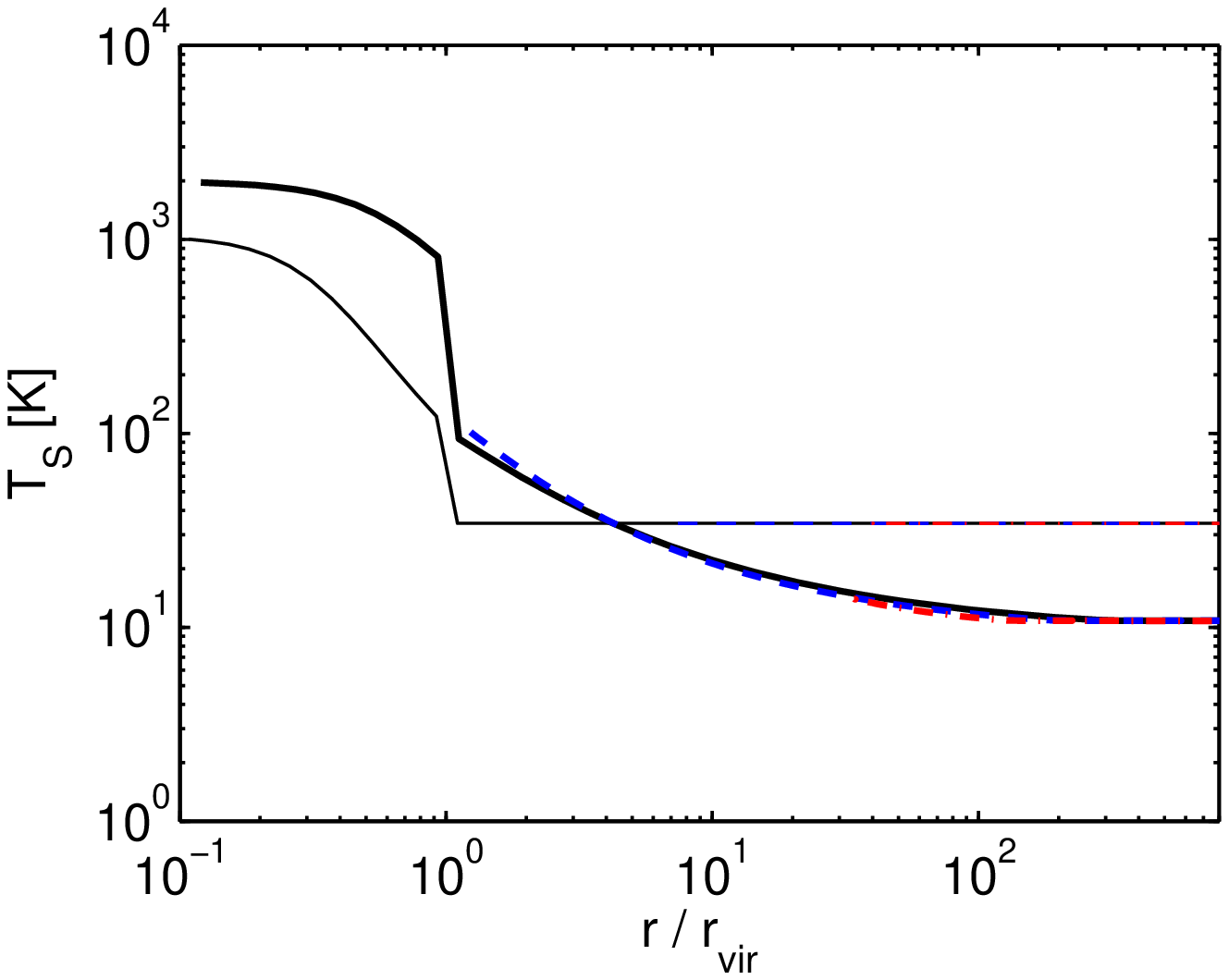}
\end{tabular}
\caption{Spin temperature profiles of minihalos/dwarf galaxies of
different masses: $M = 10^6 M_\odot$ (solid curves), $M = 10^7
M_\odot$ (dashed curves), and $M = 10^8 M_\odot$ (dot-dashed
curves), respectively. A fiducial value of $f_{\rm X}=0.1$ is
assumed for the X-ray background. The set of thick curves is for
redshift 20, and the set of thin curves is for redshift 10. {\it
Left:} the case for minihalos in CIE. {\it Right:} the case for
dwarf galaxies that are photonionized after a star burst with IMF
model-A and metallicity $Z = 10^{-7}$.} \label{Fig.Ts}
\end{figure*}

In Fig.\ref{Fig.TsTk}, we plot the spin temperature as a function of
distance from the halo center for a minihalo (left panel) and a
dwarf galaxy (right panel) of the same mass $M=10^7 M_\odot$ at
redshift $z=10$. We have taken $f_{\rm X}=0.1$ as the fiducial value
for the X-ray background. In order to show the coupling effects, the
kinetic temperature of the gas and the CMB temperature are also
plotted. As we see clearly from the left panel, the spin temperature
of the minihalo is coupled to the gas kinetic temperature ($T_{\rm
vir}$ of the halo) at the center. The halo with $M=10^7 M_\odot$ and
$z=10$ has $T_{\rm vir}\sim 5000$~K, and the gas is almost neutral
in collisional ionization equilibrium. As a result, the Ly$\alpha$
photons from recombination is totally negligible and the collisional
coupling dominates. As the radius increases, the collisional
coupling becomes less and less effective because of the decreasing
density, and the spin temperature gradually decouples from the
kinetic temperature. In addition to the dropping density, the
collisional de-excitation rate coefficients ($\kappa_{10}^{\rm eH}$
and $\kappa_{10}^{\rm H\!H}$) also decrease due to the sharply
decreasing kinetic temperature outside $r_{\rm vir}$, dramatically
reducing the collisional coupling effect. However, the Ly$\alpha$
background at $z=10$ is already strong enough to couple the spin
temperature closely to the kinetic temperature of the gas. And the
recombination in the IGM also produces a significant amount of
Ly$\alpha$ photons. As a result, $T_{\rm S}$ is always coupled to
$T_{\rm K}$ outside minihalos at $z = 10$.

As for the spin temperature profile of the dwarf galaxy in the
right panel of Fig.\ref{Fig.TsTk}, the IMF model-A and a
metallicity of $Z = 10^{-7}$ are assumed. We stop the $R_{\rm
HII}$ and $T_{\rm K}$ evolution at $t_{\rm value} = t_{\rm H} -
t_{\rm F}$, where $t_{\rm F}$ is the most probable formation time
of this halo, and the spin temperature is also evaluated at this
time. The left-hand side cut-offs of these curves are just the
position of $R_{\rm HII}$ (7.02 $r_{\rm vir}$). As the HII radius
of this galaxy is larger than the virial radius, the low density
and low temperature outside the HII region make the collisional
coupling very weak. In this case, the Ly$\alpha$ pumping is always
the dominating coupling effect. The global Ly$\alpha$ photons from
recombination and background flux dominate over the local
Ly$\alpha$ photons from soft X-ray cascading, and just as the case
outside a minihalo, the spin temperature sticks to the gas kinetic
temperature.

Then we investigate how the spin temperature changes with halo
mass and redshift, and plot two sets of curves for various halo
masses in Fig.\ref{Fig.Ts}, one (the thin set) for $z = 10$, and
the other (the thick set) for $z = 20$. A fiducial value of
$f_{\rm X}=0.1$ is assumed here. For the minihalos in the left
panel, the spin temperature is closely coupled to the kinetic
temperature outside the $r_{\rm vir}$ for halos at $z=10$, in part
because of the Ly$\alpha$ background, and in part because of the
accumulated Ly$\alpha$ photons from recombination in the IGM that
is partially ionized by the X-ray background. But at $z=20$, both
the Ly$\alpha$ background and the X-ray background are still weak,
and $T_{\rm S}$ lies between the $T_{\rm CMB}$ and $T_{\rm K}$ for
gas around minihalos at this redshift. Inside the halos, the
density is larger at higher redshift, and the collisional coupling
effect is correspondingly stronger, so the spin temperature is
more tightly coupled to the virial temperature. On the other hand,
halos are more concentrated at lower redshift, and the slope of
gas density profile is steeper. As a result the spin temperature
drops more rapidly with radius at lower redshift. One different
feature for the $10^8M_\odot$ halo is that its spin temperature is
effectively coupled to the gas kinetic temperature out to $r_{\rm
vir}$, because for this relatively high mass halo, its virial
temperature ($\sim 1.5\times 10^4\K$ for $z=10$, and $\sim
2.3\times10^4\K$ for $z=20$) is high enough to make the gas
collisionally ionized ($x_i\sim52\%$ for $z=10$, and $x_i\sim
98\%$ for $z=20$). With this partially ionized gas, Ly$\alpha$
photons from recombinations are effectively trapped, and serve as
a strong coupling agent.

\begin{figure*}
\begin{tabular}{cc}
\includegraphics[totalheight=6cm,width=7.5cm]{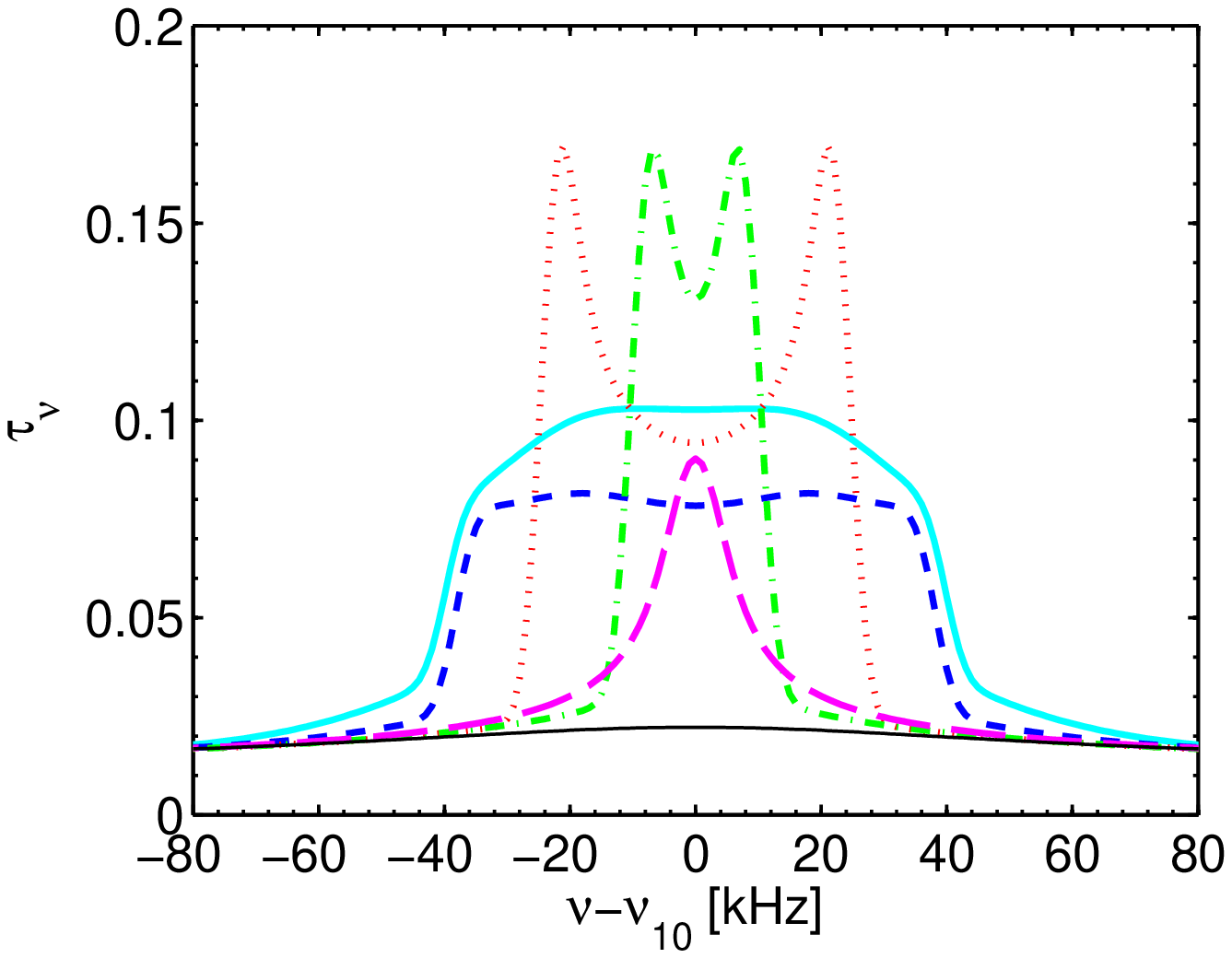}
&\includegraphics[totalheight=6cm,width=7.5cm]{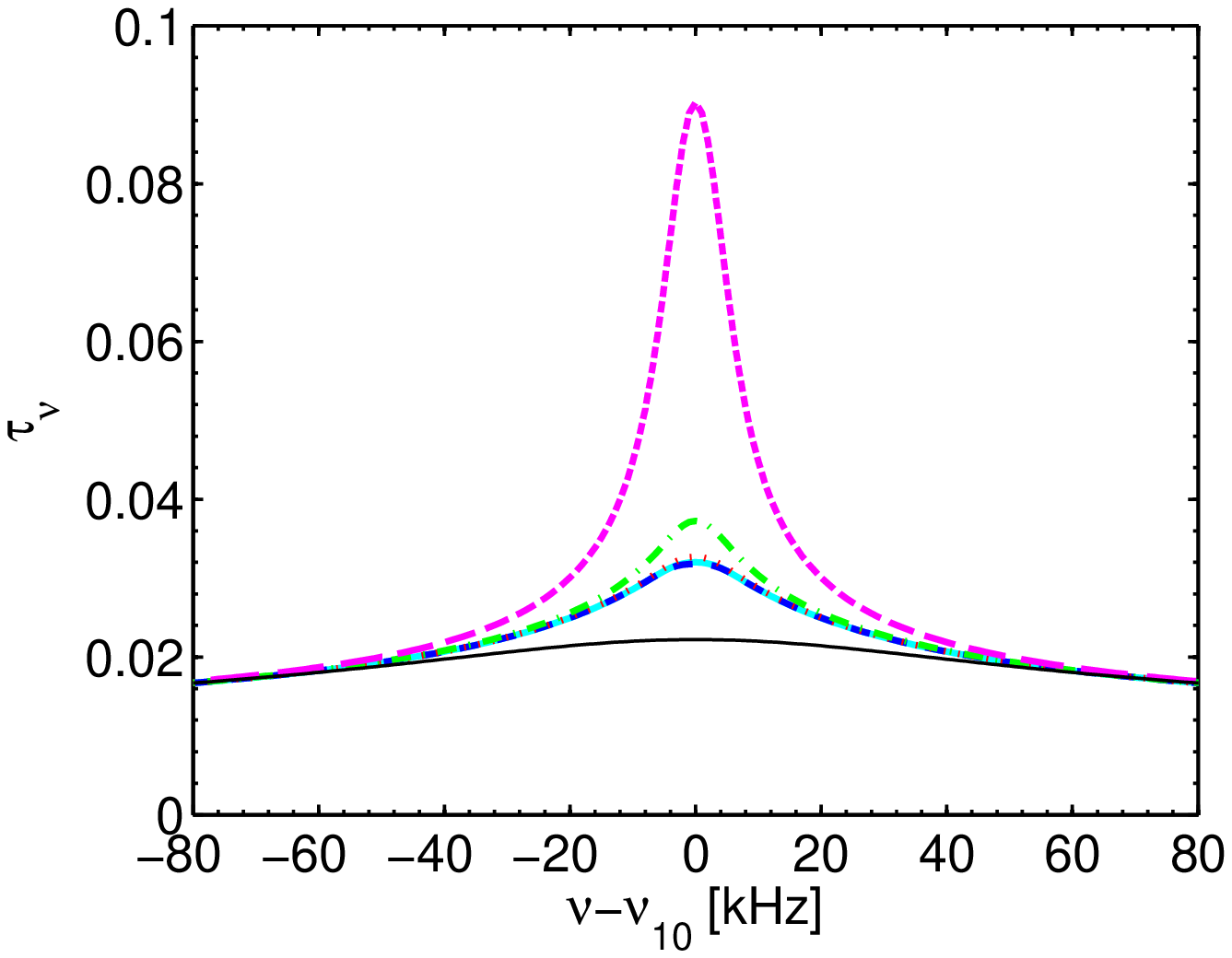}
\end{tabular}
\caption{Optical depth profiles of a minihalo/dwarf galaxy with mass
$M=10^7M_\odot$ at redshift 10. All the curves take $f_{\rm X}=0.1$.
The impact parameters of the thick curves are $\alpha = $ 0 (solid
cyan), $0.3$ (short-dashed blue), $1$ (dotted red), $3$ (dot-dashed
green), and $10$ (long-dashed magenta), respectively, and the thin
solid black curve is for $\alpha = 30$. {\it Left:} the case for a
minihalo in CIE. {\it Right:} the case for a dwarf galaxy which is
photonionized after a star burst with IMF model-A and metallicity $Z
= 10^{-7}$.} \label{Fig.depth_a}
\end{figure*}

In the right panel of Fig.\ref{Fig.Ts}, the spin temperature
profiles for dwarf galaxies are all cut off at their HII radii at the
left-hand side. For the case of $10^6 M_\odot$ dwarf galaxy, due to the
much lower star formation efficiency
($f_\star \sim 4\times10^{-5}$, which increases to
$\sim 0.03$ for a galaxy with $10^8 M_\odot$) \citep{SF09},
it could only create a very small HII
region ($R_{\rm HII}\sim 0.008\, r_{\rm vir}$ for $z=10$, and
$R_{\rm HII}\sim 0.02\, r_{\rm vir}$ for $z=20$).
As for the galaxies with $10^7 M_\odot$ and $10^8 M_\odot$, they
have larger stellar masses and create larger HII regions. So they
have much less gas which could contribute to the 21 cm line
absorption.

\subsection{The Absorption Line Profiles}

Consider an isolated minihalo with mass $M=10^7M_\odot$ at redshift
$z=10$ for which we compute the optical depth as a function of
frequency using Eq.(\ref{Eq.tau}). Results are shown in the left
panel of Fig.\ref{Fig.depth_a} for different impact parameters of
the lines of sight. Very interesting features can be seen in the
shown profiles. First, there are two peaks which sandwiched the
center of the line for impact parameters $\alpha=1$ and $\alpha=3$,
showing horn-like profiles; second, as the impact parameter
increases, the peak optical depth is NOT always decreasing. These
two points are related issues, and they are both caused by the
infalling gas with low spin temperature outside the minihalo.

\begin{figure}
\centering{\resizebox{8.5cm}{8cm}{\includegraphics{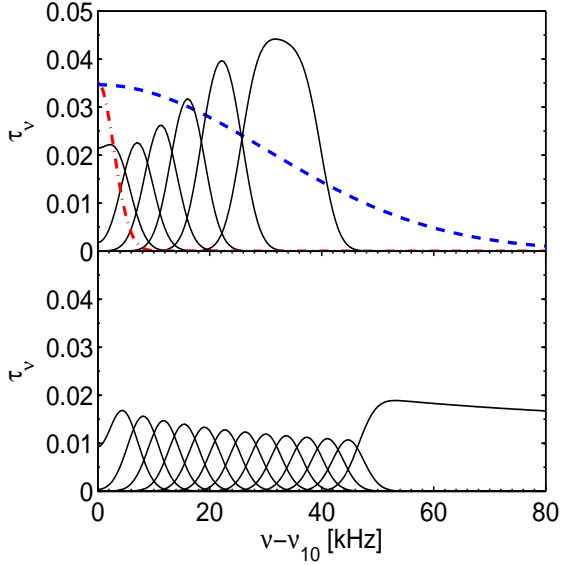}}}
\caption{Different contributions to the optical depth from the gas
at different radii. {\it Upper panel:} the absorption by gas inside
of $r_{\rm vir}$ is shown as the dashed line, the solid lines from
right to left correspond to the absorptions by segments of $(1-2)\,
r_{\rm vir}$, $(2-3)\, r_{\rm vir}$, $(3-4)\, r_{\rm vir}$, $(4-5)\,
r_{\rm vir}$, $(5-6)\, r_{\rm vir}$, and $(6-7)\, r_{\rm vir}$,
respectively, and the dotted-dashed line represents the absorption
by the segment of $(7-8)\, r_{\rm vir}$. {\it Bottom panel:} the 12
solid lines from left to right correspond to the absorptions by
segments of $1\, r_{\rm vir}$ each starting from $8\, r_{\rm vir}$,
and the last curve represents the integral absorption from 20 to 100
$r_{\rm vir}$.} \label{Fig.segments}
\end{figure}

To see clearly the origin of these interesting profiles, we have to
understand the different contributions to the 21 cm absorption from
the gas located at different radii. Because of the infall, the
absorption line produced by the gas at the far side of the halo is
blue-shifted, while that produced by the gas at the near side of the halo
is redshifted. Let $\nu_p(r)$ be the peak frequency of
optical depth created by gas located at radius $r$, and according to
the integrand in Eq.(\ref{Eq.tau}), the absorption is shifted
from the line center by
\begin{equation}
\nu_p(r) - \nu_{\rm 10} \,=\, \frac{\bar v(r)}{c}\, \nu_{\rm 10}.
\end{equation}
Considering a line of sight passing through a minihalo with
$M=10^7M_\odot$ and $z=10$ from the center ($\alpha =0$), we divide
the integration into segments, each of which has a contribution from
a length of $1\, r_{\rm vir}$ on the line of sight. Then we plot
every absorption line created by one segment in
Fig.\ref{Fig.segments}. In the upper panel, the absorption line
produced by the gas inside of $r_{\rm vir}$ is shown as the dashed
curve, the solid lines from right to left correspond to the
absorptions by segments of $(1-2)\, r_{\rm vir}$, $(2-3)\, r_{\rm
vir}$, $(3-4)\, r_{\rm vir}$, $(4-5)\, r_{\rm vir}$, $(5-6)\, r_{\rm
vir}$, and $(6-7)\, r_{\rm vir}$, respectively, and the
dotted-dashed line represents the absorption by the segment of
$(7-8)\, r_{\rm vir}$. In the bottom panel, the 12 solid lines from
left to right correspond to the absorptions by segments of $1\,
r_{\rm vir}$ each starting from $8\, r_{\rm vir}$, and the last
curve on the right represents the integral absorption from 20 to 100
$r_{\rm vir}$.

As seen clearly from Fig.\ref{Fig.segments}, the virialized gas
inside the minihalo has no bulk velocity, and the peak optical depth
is located at the center. The gas at $(1-2)\, r_{\rm vir}$ has the
highest infall velocity, and the corresponding profile lies at the
largest distance to the line center in the upper panel. As the
radius increases, the $\tau_\nu$ profile gets closer to the line
center because of the lower infall velocity, and the optical depth
decreases slowly with the decreasing density. One interesting
feature is that there is a special position where the total velocity
of the gas changes from negative (infall dominated) to positive
(Hubble flow dominated), and at this turning point the two
absorption lines created by the two segments on both sides of the
minihalo converge into one at the line center, and they contribute
substantially to the central optical depth. For the minihalo with
$M=10^7M_\odot$ and $z=10$, this turning point lies at $7.3\, r_{\rm
vir}$, and consequently the absorption by the segment of $(7-8)\,
r_{\rm vir}$ peaks at the line center. After that, the $\tau_\nu$
profiles that come from larger radii leave the line center again,
because the infall velocity becomes even smaller and the total
velocity (in the same direction as Hubble flow) becomes more and
more positive. Finally, when it goes out of the region influenced by
the minihalo's gravity, and the density drops to the cosmic mean
value, we recover the IGM optical depth.

The line profiles in Fig.\ref{Fig.depth_a} are better understood by
noting that each is an integral of contributions from different
radii. A substantial contribution to the optical depth comes from
the outer region, because the gas outside $r_{\rm vir}$  has lower
spin temperature. Especially, the gas in the infalling region on the
far (near) side of the halo shares the same bulk velocity with the
gas out of the region on the near (far) side, and they absorb the 21
cm photons at the same frequency. Therefore, the optical depth at
the frequency range corresponding to the infalling region is
increased a lot. In addition, the line profile gets narrower with
lower temperature compared to the gas inside the halo, which further
increases the peak value. As the impact parameter $\alpha$ increases
from 0 to 1, more contribution comes from this cold region, so the
peak optical depth increases. Also, the infall velocity shifts the
peaks away from the center, and results in the horn-like profile. As
$\alpha$ increases further up to $3$, the infall velocity decreases,
and the two peaks move closer to each other. For $\alpha$ larger
than the radius of the velocity turning point, the two peaks merge
together and decrease slowly to the IGM optical depth as $\alpha \to
\infty$.

The 21 cm profiles for an isolated dwarf galaxy for various impact
parameters are shown in the right panel of Fig.\ref{Fig.depth_a}.
Just as before, the spin temperature and the optical depth profiles
are all evaluated at the time $t_{\rm value} =t_{\rm H}-t_{\rm F}$,
and the most probable value for $z_{\rm F}$ is used. Thus we get an
upper limit on the ionization, heating, and Ly$\alpha$ coupling
effects. The line profiles are completely different from the case of
minihalos. The horn-like profiles disappears and the absorption is
strongly reduced for small impact parameters.
The dwarf galaxy with $M=10^7 M_\odot$ and $z=10$ has an HII radius
of $7.3\,r_{\rm vir}$, which is close to the turning point of the
gas velocity. In other words, the hydrogen atoms inside $r_{\rm
vir}$ and those within the infalling region are totally ionized,
erasing the absorption features contributed by hydrogen in this
region. Therefore, there will be only one peak at the center no
matter what the impact parameter is, and the optical depth is
reduced for lines of sight which penetrate the HII region. As the
impact parameter $\alpha$ increases, the optical depth first
increases because more neutral gas near the HII region is
intercepted by the line of sight. It reaches a maximum when $\alpha
\approx R_{\rm HII}$ and then drops, approaching the IGM optical
depth.

\begin{figure*}
\begin{tabular}{cc}
\includegraphics[totalheight=6cm,width=7.5cm]{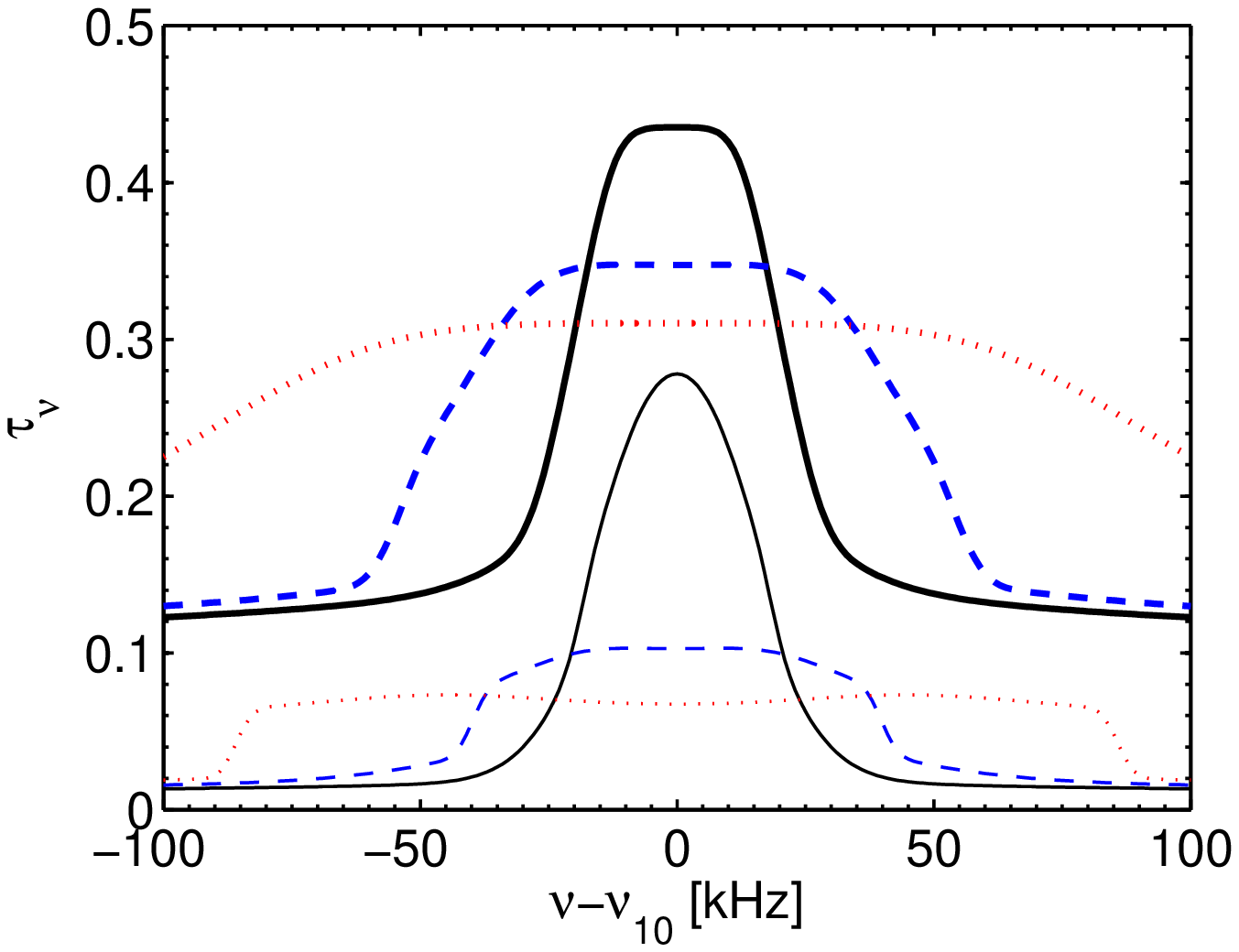}
&\includegraphics[totalheight=6cm,width=7.5cm]{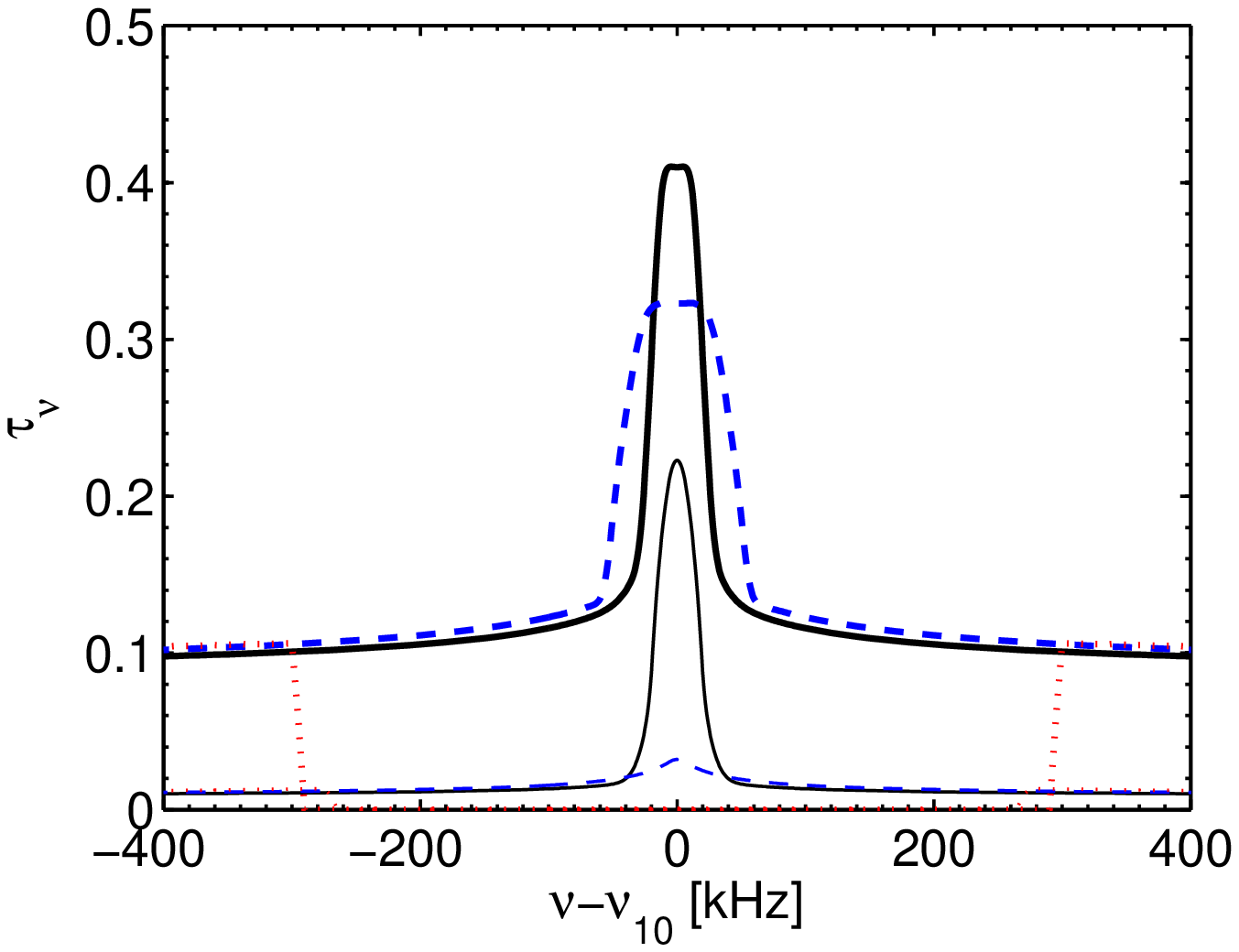}
\end{tabular}
\caption{Optical depth profiles of minihalos/dwarf galaxies with
different masses: $M = 10^6 M_\odot$ (solid curves), $M = 10^7
M_\odot$ (dashed curves), and $M = 10^8 M_\odot$ (dotted curves),
respectively. The thick curves are for redshift 20, while the thin
curves are for redshift 10. All the curves take $f_{\rm X}=0.1$.
{\it Left:} the case for minihalos in CIE. {\it Right:} the case for
dwarf galaxies that are photonionized after a star burst with IMF
model-A and metallicity $Z = 10^{-7}$. The impact parameter shown
here is $\alpha = 0$.} \label{Fig.depth_zM}
\end{figure*}

We also show the optical depth profiles for different halo masses
and redshifts in Fig.\ref{Fig.depth_zM}. Here we also take $f_{\rm
X}=0.1$ for the X-ray background. In the left panel, the case for
minihalos is shown. As we expected, the line profiles are broader
for halos with greater masses. That is because, on one hand, the
virial temperature is higher for halos of greater masses, which
results in a broader Doppler profile for the absorption by gas
inside the halo, and on the other hand, more massive halos have
stronger gravitational influence on the surrounding gas, and the
induced higher infalling velocity shifts the absorption line farther
from the line center.
In general, the absorption is stronger at higher redshift. The
reason is that the IGM is denser at higher redshift, and also, the
X-ray background is gradually set up as redshift decreases, and it
heats and partially ionizes the gas in the IGM. As the frequency
gets far from the line center, the optical depths of all the
minihalos approach to the mean IGM value.

In the right panel of Fig.\ref{Fig.depth_zM}, we show the 21 cm line
profiles for dwarf galaxies with the same masses and redshifts as
those for minihalos in the left panel. Star bursts with IMF model-A
and metallicity $Z=10^{-7}$ are assumed. An interesting feature
emerges for the galaxy with $10^8 M_\odot$. Because of the higher
star formation efficiency associated with higher mass galaxies, this
dwarf galaxy creates a large HII region ($R_{\rm HII}\sim 38\,
r_{\rm vir}$ for $z=10$, and $R_{\rm HII}\sim 34\, r_{\rm vir}$ for
$z=20$) erasing all the absorption inside of it. As a result, a
broad optical depth trough is produced instead of an absorption
line! In the case of IMF model-C, the dwarf galaxy of the same mass
could ionize an even larger HII region and hence could result in an
even broader optical depth trough.

\begin{figure}
\centering{\resizebox{8cm}{6cm}{\includegraphics{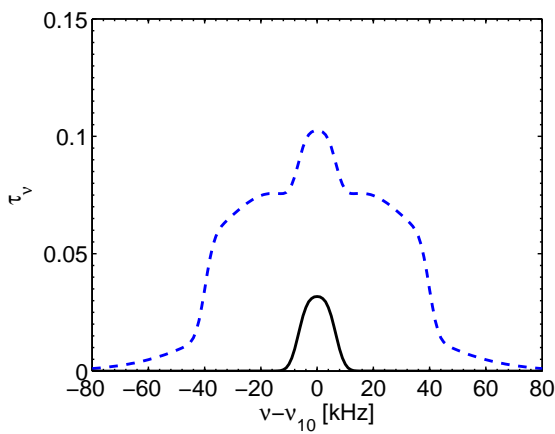}}}
\caption{The optical depth profiles of a minihalo (blue dashed
curve) and a dwarf galaxy (black solid curve) with $M=10^7M_\odot$
at $z=10$. A star burst with IMF model-A and metallicity $Z =
10^{-7}$ is assumed for this galaxy. The optical depth at each point
is integrated up to $D_{\rm min}/2$, the impact parameter is $\alpha
= 0$, and the X-ray parameter is $f_{\rm X}=0.1$.}
\label{Fig.depth2Dmin}
\end{figure}

All the line profiles above are computed assuming that the minihalo
or the dwarf galaxy is isolated. In real cosmic structures, a halo
is surrounded by other halos, and if we integrate the optical depth
to a distance larger than the mean separation $D$ of the halos, we
will probably hit another halo. So the integration should only be
considered as reliable up to a distance of $D/2$. In practice, a
$10^8 M_\odot$ halo, for example, might have many smaller halos
closer to it than another $10^8 M_\odot$ halo. Therefore, it is the
mean separation of the smallest halos we are considering, i.e. the
halos with $M=10^6 M_\odot$ (for $f_{\rm X}\lesssim 0.2$), that
determines the integration limit. We denote this mean separation as
$D_{\rm min}$. The maximum impact parameter of a line of sight
should also be $\alpha_{\rm max} = D_{\rm min}/2$. Integrating the
optical depth up to $D_{\rm min}/2$, we plot optical depth profiles
of a minihalo (blue dashed curve) and a dwarf galaxy (black solid
curve) with $M=10^7M_\odot$ at $z=10$ in Fig.\ref{Fig.depth2Dmin}.
The line of sight is assumed to be passing through the
minihalo/dwarf galaxy from the center, and we set $f_{\rm X}=0.1$.
We see that the dwarf galaxy only produces a narrow and weak
absorption line at the center, because for this galaxy, only the gas
in a sphere between the HII radius ($\sim 7.3\, r_{\rm vir}$) and
$D_{\rm min}/2$ ($\sim 9.1\, r_{\rm vir}$) contributes to the
absorption. In reality, however, the optical depth will not drop to
zero at the boundaries of the absorption line but will connect with
another line created by a neighboring halo. In addition, as we
mentioned before, the clustering of dwarf galaxies will extend their
HII regions. Similarly, some minihalos will be clustered around the
dwarfs, and the surrounding gas in the infall region may be ionized,
even if the minihalos themselves can self-shield. Therefore, the
clustering can reduce the 21 cm optical depth of some minihalos and
dwarf galaxies. Although the problem should ideally include the
clustering properties of early galaxies, it is beyond the scope of
this paper to include such features.

\section{THEORETICAL SPECTRUM}\label{theoSpec}

\begin{figure}
\centering{\resizebox{8cm}{6cm}{\includegraphics{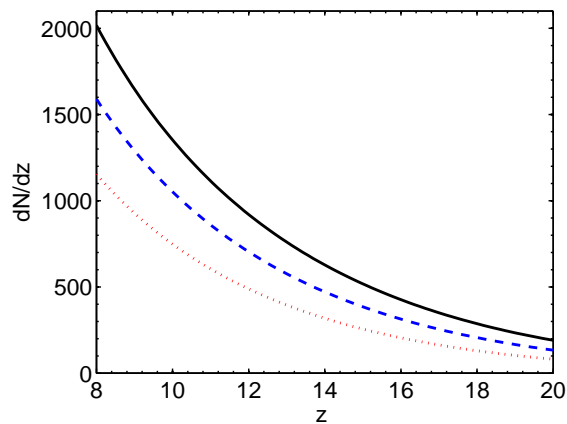}}}
\caption{Theoretical evolution of line number density produced by
minihalos/dwarf galaxies along a line of sight. The maximum halo
mass is $10^{10}M_\odot$, and the minimum halo masses are
$10^6M_\odot$ (solid), $2 \times 10^6 M_\odot$ (dashed) and $5
\times 10^6 M_\odot$ (dotted), respectively.} \label{Fig.dNdz}
\end{figure}

In order to superimpose the 21 cm absorption lines of minihalos and
dwarf galaxies onto a radio spectrum of GRB afterglow, we first have
to compute the line number density per unit redshift. Theoretically,
including all (weak and strong) absorption lines, the number of halo
intersections along a line of sight per redshift interval is
\begin{equation}
\frac{dN}{dz} \,=\, (1+z)^2\, \frac{dr}{dz}\, \int_{M_{\rm
min}}^{M_{\rm max}} n(M,z)\, A_{\rm max}\, dM,
\end{equation}
where $dr/dz$ is the comoving radial distance per redshift interval,
$n(M,z)$ is the halo mass function given by Eq.(\ref{Eq.haloMF}),
and $A_{\rm max}=\pi\alpha_{\rm max}^2$ is the cross-section (in
physical coordinates) of a halo, in which the maximum impact
parameter $\alpha_{\rm max}$ is set by half of the mean halo
separation, i.e. $D_{\rm min}/2$, at each redshift. We plot the line
number density as a function of redshift in Fig.\ref{Fig.dNdz} for
three different values of $M_{\rm min}$, which are appropriate for
different levels of the X-ray background.
The curves are cut off at redshift $z = 8$ since our model applies
only to the early stages of reionization, when the stellar sources
have not set up an ionizing background.

This line number density is also the probability $P(z)$ that a line
of sight intersects an object in a redshift interval $dz$ centered
on $z$. In analogy to the method used in \citet{Furlanetto02}, we
divide the observed frequency band into small bins, each of which
corresponds to a redshift interval $\Delta z$ that is small enough
to make sure that the probability to have an intersection in each
bin $P(z)\,\Delta z < 0.01$. Then we generate a random number $R_i$
(uniformly distributed in $[0, 1)$) for each bin, and an
intersection is said to take place if the condition $R_i <
P(z)\,\Delta z$ is satisfied.

\begin{figure}
\centering{\resizebox{9cm}{9cm}{\includegraphics{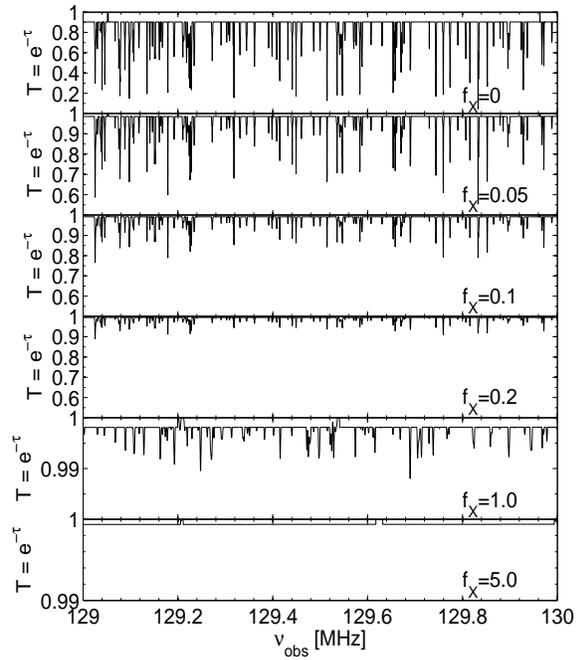}}}
\caption{Relative transmission along a line of sight at redshift
$z\sim 10$. The six panels from top to bottom show the results for
$f_{\rm X}=0$, $f_{\rm X}=0.05$, $f_{\rm X}=0.1$, $f_{\rm X}=0.2$,
$f_{\rm X}=1.0$, and $f_{\rm X}=5.0$, respectively. Note that the
y-axes are different between panels.} \label{Fig.trans}
\end{figure}

When an intersection occurs, we randomly choose a mass $M$ for the
intersected halo 
according to the mass function. With the halo mass and the redshift
of that bin, we compute its formation redshift distribution, and
$z_{\rm F}$ is Monte-Carlo sampled from the distribution curve. Then
we use the star formation criterion described in section
\ref{SBcriterion} to determine whether it is a minihalo in
collisional ionization equilibrium or a dwarf galaxy photonionized
after a star burst. For every dwarf galaxy, its metallicity is
fitted as a function of halo mass to the results given by
\citet{SF09}. As the fitted metallicity is always $\gtrsim 10^{-4}$,
the IMF model-A is used. With the impact parameter randomly selected
with an equal probability per unit area for each object, we compute
the 21 cm line profile for every minihalo and dwarf galaxy
intersected by a sightline. From $129 \MHz$ ($z\sim 10$) to $158
\MHz$ ($z\sim 8$), we found 3241 lines if we set $M_{\rm min} = 10^6
M_{\rm \odot}$, out of which 210 lines are attributed to dwarf
galaxies. That is about 6.5\% of the lines coming from dwarf
galaxies, and the rest come from minihalos. The relative
transmission $T=\exp(-\,\tau)$ for a range of observed frequency
corresponding to $z\sim 10$ is shown in Fig.\ref{Fig.trans}. The six
panels from top to bottom show the results for $f_{\rm X}=0$,
$0.05$, $0.1$, $0.2$, $1$ and $5$, respectively. $M_{\rm min} = 10^6
M_{\rm \odot}$ is used for $f_{\rm X} \le 0.2$, $M_{\rm min} =
2\times 10^6 M_{\rm \odot}$ is used for $f_{\rm X}=1$, and we use
$M_{\rm min} = 5\times 10^6 M_{\rm \odot}$ for the case of $f_{\rm
X}=5$. Note that the y-axes are different between panels.

As seen from the figure, the absorption lines are very narrow and
closely spaced, resembling a 21 cm forest. The depth of the 21 cm
absorption strongly depends on the existence and the intensity of
the uncertain X-ray background. The absorption strength is
significantly increased if there is no X-ray background, and it
decreases rapidly with the increase of the X-ray background, which
ionizes and heats the gas in the IGM reducing the optical depth of
both non-linear structures and the global absorption. Especially, if
the X-ray background in the early universe was very strong (e.g.
$f_{\rm X}=5$), resulted from a large number of high mass X-ray
binaries in the case of top-heavy IMF towards high-$z$ or from
mini-quasars, we could hardly see any feature on a spectrum. This is
because, on one hand, the line number density is reduced, and on the
other hand, the absorption by non-linear structures is so weak that
most lines are covered by the global absorption. For some minihalos,
the gas outside $r_{\rm vir}$ contributes to the absorption far from
the line center because of the infall velocity, and a few absorption
lines of the neighboring halos overlap with each other on the wings.
In this case the actual optical depth of each point is the sum over
all the absorptions caused by these overlapping lines.

We compared two spectra: one taking into account the local X-rays
contributing to heating and cascading to Ly$\alpha$ photons, the
second neglecting the above. It turns out that for the IMF model-A
and the relatively high metallicities that are used in the spectrum,
the small amount of soft X-rays produced by the dwarf galaxies make
little difference (less than 0.1\%) even without an X-ray
background. So we could not see any signature that can be attributed
to them in the spectrum. However, if a dwarf galaxy with relatively
high mass and (almost) primordial composition does exist, a
top-heavy IMF (like the IMF model-C) is expected
\citep{Schneider02,Schneider03},
and the associated large HII region will result in a broad bump on
the spectrum, erasing several absorption lines produced by
neighboring minihalos.

\begin{figure}
\centering{\resizebox{7.5cm}{6cm}{\includegraphics{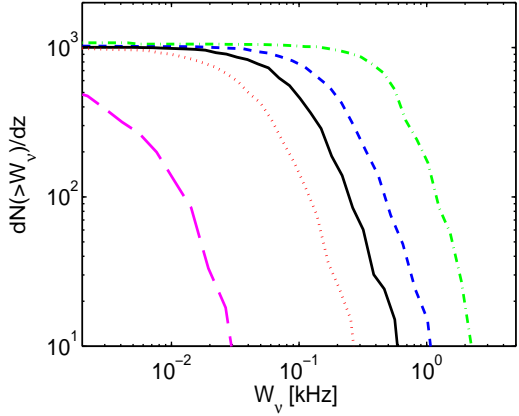}}}
\caption{Cumulative distribution of equivalent width of the 21 cm
absorption lines around $z \sim 10$. The dot-dashed, short dashed,
solid, dotted, and long dashed curves are computed assuming $f_{\rm
X}=0$, $f_{\rm X}=0.05$, $f_{\rm X}=0.1$, $f_{\rm X}=0.2$, and
$f_{\rm X}=1.0$, respectively.} \label{Fig.EWdist}
\end{figure}

Directly from the spectrum, we could compute the distribution of
equivalent width of the absorption lines for a specific range of
observed frequency corresponding to a specific redshift. As the
continuum of a background source has a global decrement due to the
absorption of the diffuse IGM, the real signal of non-linear
structures is the extra absorption with respect to the flux
transmitted through the IGM. Therefore, the equivalent width of an
absorption line should be defined as
\begin{eqnarray}
W_\nu &=& \int\, \frac{\displaystyle f_c\, e^{-\,\tau_{\rm IGM}(z)}
\,-\, f_c\, e^{-\,\tau(\nu)}}{\displaystyle f_c\, e^{-\,\tau_{\rm IGM}(z)}}\; d\nu \nonumber \\
&=& \int\, {\displaystyle (1 \,-\, e^{\tau_{\rm IGM}(z) \,-\,
\tau(\nu)})}\; d\nu,
\end{eqnarray}
where $f_c$ is the continuum flux of the background radio source,
and $\tau_{\rm IGM}(z)$ is the optical depth of the diffuse IGM at
redshift $z$. Using the theoretical spectrum of 129 -- 133 MHz
(corresponds to $z=10.01 - 9.68$), we compute the cumulative
distribution of equivalent width of those 21 cm lines around $z\sim
10$, which is shown in Fig.\ref{Fig.EWdist}. The dot-dashed, short
dashed, solid, dotted, and long dashed curves take $f_{\rm X}=0$,
$0.05$, $0.1$, $0.2$, and $1$, respectively. As for the case of
$f_{\rm X}=5$, the number of lines emerged from the IGM absorption
in this frequency range is too small to derive any statistical
implication, so the distribution of equivalent width is not shown
for this case. For the fiducial value $f_{\rm X}=0.1$, the majority
of absorption lines have equivalent widths around $0.03$ to $0.3
\kHz$. We see that the number of large signals is very sensitive to
the presence and intensity of the X-ray background. If there is no
X-ray background ($f_{\rm X}=0$), then about $95\%$ of absorption
lines have $W_\nu > 0.1 \kHz$. For the value $f_{\rm X}=0.05$, there
are about $75\%$ of lines with $W_\nu
> 0.1 \kHz$, while for $f_{\rm X}=0.1$ and $0.2$, this fraction
drops to $45\%$ and $15\%$ respectively. If $f_{\rm X}=1$ or even
higher, there will be no absorption line gets $W_\nu > 0.1 \kHz$.


\section{THE OBSERVABILITY}\label{obsSpec}
There are two kinds of radio sources which are potentially usable
for absorption line studies at high redshifts during the epoch of
reionization. One is the high redshift quasars, which are quite
luminous, but so far no quasar has been confirmed at $z>6.5$. The
other candidate is the radio afterglows of GRBs. Although they are
fainter than quasars, they are more likely to exist at higher
redshifts. Also, they have simpler power-law spectra at low
frequencies due to synchrotron self-absorption, so it may be easier
to extract absorption signals from bright GRB afterglows. Here we
examine the observability of 21 cm signals on both spectra of these
background sources.

As for the upcoming and planned low frequency interferometers, a
spectral resolution of $1\kHz$ is achievable for LOFAR\footnote{
http://www.lofar.org/index.htm} and
SKA\footnote{http://www.skatelescope.org/}. From the theoretical
spectrum computed above, we get about one absorption line in every
$8.4 \kHz$ at $z\sim10$ on average when we use $M_{\rm min}=10^6
M_\odot$ (line overlapping is accounted), and the line density (per
observed frequency interval) is lower for lower redshifts or higher
$M_{\rm min}$.
On the other hand, the line width ranges mostly from $\sim 1 \kHz$
to $\sim 5 \kHz$ for halos of different masses. So the instruments
can marginally resolve these 21 cm lines. While resolving the
detailed line profile is probably out of reach, the line counting is
feasible as long as sufficiently bright radio sources can be found
at high redshift.

\begin{figure*}
\begin{tabular}{cc}
\includegraphics[totalheight=6cm,width=7.5cm]{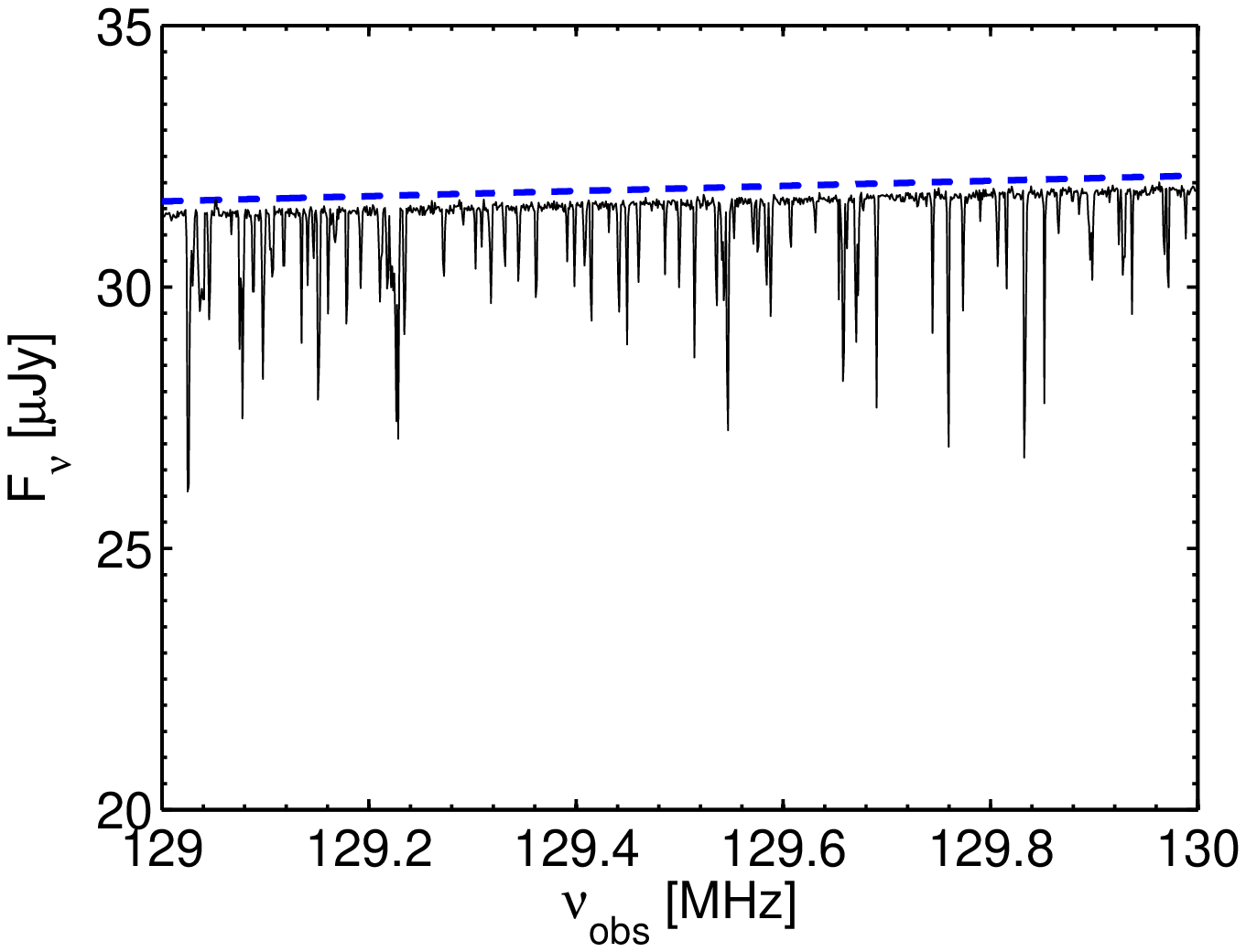}
&\includegraphics[totalheight=6cm,width=7.5cm]{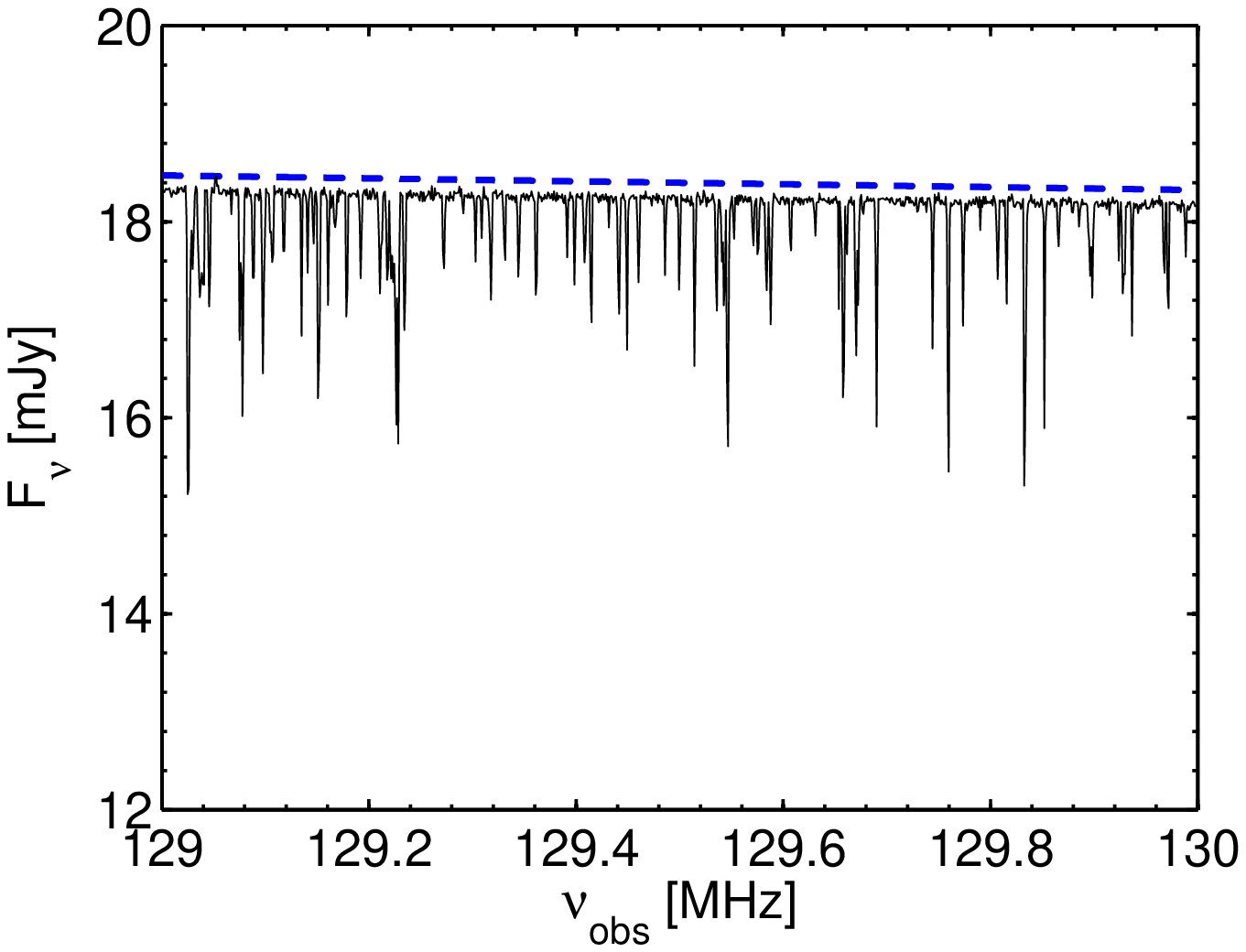}
\end{tabular}
\caption{Synthetic spectra of 21 cm absorptions against a GRB
afterglow (left panel) and a quasar (right panel) with $1\kHz$
resolution. The corresponding redshift is $z\sim 10$, and the
unabsorbed continua are shown as the dashed (blue) lines. $f_{\rm
X}=0.1$ is assumed for both spectra.} \label{Fig.Spec1kHz}
\end{figure*}

To produce mock spectra matching real observations, we degenerate
the theoretical spectrum to a resolution of $1\kHz$, add Gaussian
noise onto each pixel with the signal-to-noise ratio $S/N = 5$, and
convolve it with a continuum of GRB afterglow as well as a radio
spectrum of quasar. Two synthetic spectra are illustrated in
Fig.\ref{Fig.Spec1kHz} with the left panel for a GRB afterglow and
the right panel for a quasar. We take $f_{\rm X}=0.1$ for both
cases. The flux density of GRB afterglow is scaled to $100 \mu\Jy$
at $200\MHz$ and $z=6$, which is achievable for the afterglow of an
energetic GRB produced by explosion of a massive metal-free star at
high redshift, with isotropic energy of $10^{54} \erg$
\citep{Ioka05}, and the spectral index is taken to be $2$ ($F_\nu
\propto \nu^2$), which is appropriate for the synchrotron
self-absorption spectrum at the frequencies of interest
\citep{Frail03}. The quasar flux density is scaled to $20 {\rm m}
\Jy$ at $120\MHz$ and $z=10$, with a spectral index of $-1.05$ as
fitted to the radio spectrum of the powerful radio galaxy Cygnus A
\citep{Carilli02}.

Apart from spectral resolution considerations, we also need
high enough sensitivity to observe the absorption lines. In
other words, with the planned instruments, the background source has
to be bright enough to get the decrement of flux density higher than
the detection limit. The minimum detectable flux density of an
interferometer is related to the system temperature $T_{\rm sys}$,
the effective aperture area $A_{\rm eff}$, channel width
$\Delta\nu_{\rm ch}$, integration time $t_{\rm int}$, and the
signal-to-noise ratio $S/N$ by
\begin{equation}
\Delta F_{\rm min}\, =\, \frac{2\, k_{\rm B}\, T_{\rm sys}}{A_{\rm
eff}\sqrt{\Delta\nu_{\rm ch}\,t_{\rm int}}}\, \frac{S}{N}.
\end{equation}
The real signals of minihalos or dwarf galaxies are their additional
absorptions with respect to the absorption by the IGM, i.e. $\Delta
F = F_\nu \exp(-\tau_{\rm IGM}) - F_\nu \exp(-\tau)$. Equating this
flux decrement to the detection limit, we get the minimum flux
density of the background source required to observe the absorption
lines:
\begin{eqnarray}
F_{\rm min}&=&542 \mu\Jy \left(\frac{S/N}{5}\right)\left(
\frac{0.1}{\displaystyle e^{-\tau_{\rm IGM}} - e^{-\tau}}\right)
\left(\frac{1 \kHz}{\Delta\nu_{\rm ch}}\right)^{1/2}
\nonumber\\
&&\times \left(\frac{5000\,\m^2\K^{-1}}{A_{\rm eff}/T_{\rm
sys}}\right) \left(\frac{30\, {\rm days}}{t_{\rm int}}\right)^{1/2},
\end{eqnarray}
where the ratio $A_{\rm eff}/T_{\rm sys}$ is an intrinsic parameter
describing the sensitivity of an interferometry array, and we use
the value for SKA here. For the GRB afterglow, the integration time
is limited by its fading time scale. Typically, after a bright,
short-lived radio ``flare'' at early times, the subsequent evolution
of the radio afterglow can be described by a slow rise to maximum,
followed by several segments of power-law decays with a timescale of
$\sim 100$ days \citep{Frail03}. Here we have assumed a reasonable
integration time of 30 days, and find that a minimum flux density of
$\sim 500\, \mu\Jy$ is required to detect the absorption lines with
the resolution of $1\kHz$ for the case of $f_{\rm X}=0.1$.

As seen from the continua (dashed lines) in Fig.\ref{Fig.Spec1kHz},
the flux density of our prototype GRB afterglow is more than one order of magnitude 
lower than this limit. Note that this is already an energetic GRB
which is one order of magnitude brighter than normal GRBs. It may be
possible that there are even brighter GRBs, but for most GRBs it
seems that the radio afterglows would be too dim for being used as
background source in observations with such a high spectral
resolution. If we by chance find a quasar at very high redshift
during the early stages of reionization, the signals could be easily
detected. Especially, if one stacks together several lines to get an
average profile, it will hopefully reveal the horn-like profiles we
found.

Alternatively, we may try  broadband observations with lower
resolution. In this case, we could use the standard measurement of
$D_{\rm A}$, i.e. the mean (relative) flux decrement in each band,
in analogy to the Ly$\alpha$ forest experiments of quasars
\citep{Rauch98}. We redefine the mean flux decrement with respect to
the continuum after the absorption by the IGM:
\begin{equation}
D_{\rm A} \,=\, \langle \frac{f_{\rm IGM} - f_{\rm obs}}{f_{\rm
IGM}}\rangle \,=\, \langle 1 - e^{\tau_{\rm IGM} - \tau}\rangle
\,=\, 1 - e^{\tau_{\rm IGM} - \tau_{\rm eff}},
\end{equation}
where the angular brackets represent the average over each band and
$\tau_{\rm eff}$ is the effective optical depth of the band. $D_{\rm
A}$ measures the excessive absorption by minihalos/dwarf galaxies
compared to the diffuse IGM, and the different values of $D_{\rm A}$
measured in different bands represent the absorptions at different
redshifts, showing the evolution of non-linear structures during the
epoch of reionization. In terms of $D_{\rm A}$, the flux decrement
can be written as $\Delta F = F_\nu \exp(-\tau_{\rm IGM}) - F_\nu
\exp(-\tau_{\rm eff}) = F_\nu D_{\rm A}\exp(-\tau_{\rm IGM})$, then
the requirement of the background source is
\begin{eqnarray}
F_{\rm min}&=&77.4 \mu\Jy \left(\frac{S/N}{5}\right)\left(
\frac{0.99}{\displaystyle e^{-\tau_{\rm IGM}}}\right) \left(
\frac{0.01}{D_{\rm A}}\right)
\nonumber\\
&&\times \left(\frac{5 \MHz}{\Delta\nu_{\rm ch}}\right)^{1/2}
\left(\frac{5000\,\m^2\K^{-1}}{A_{\rm eff}/T_{\rm sys}}\right)
\left(\frac{30\, {\rm days}}{t_{\rm int}}\right)^{1/2}.
\end{eqnarray}

With the broadband observation, the major concern is that the
available length of the line of sight is limited. On one hand, we
wish to put the background source as far as possible in order to get
enough bands to see varying $D_{\rm A}$ values at different
redshifts. On the other hand, as the source moves farther, it gets
dimmer, lower the average signal to noise ratio, then an even
broader band is required to detect the signal, thus beyond a certain
distance the number of useful bands might decrease again. An optimal
source redshift $z_{\rm GRB}$ can be found to maximize the number of
available bands. Considering the applicability of our model, we set
the lower limit of the redshift to be $z_{\rm lim}=8$, and find that
the optimal redshift for the GRB is $z_{\rm GRB} \sim 9.8$. This
redshift depends mainly on the lower limit of the redshift we set,
but of course it is also dependent somewhat on the reionization
history, e.g. $D_{\rm A}$ which evolves slowly with redshift. This
does not depend on the observational parameters.
The optimal redshift is quite encouraging given that we have already
seen 2 GRBs beyond $z = 6$.

\begin{figure}
\centering{\resizebox{9cm}{9cm}{\includegraphics{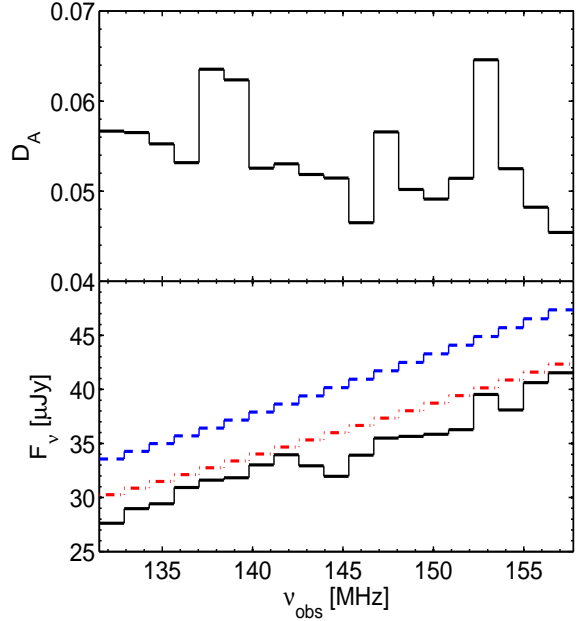}}}
\caption{The mean flux decrement with respect to the continuum
absorbed by the diffuse IGM ({\it upper panel}) and the synthetic
broad band spectrum with $1.38\,\MHz$ resolution ({\it bottom
panel}) in the case of $f_{\rm X}=0$. The 19 pixels show the
frequency range corresponding to the redshift range from 9.8 to 8.
In the {\it bottom panel}, the dashed, dot-dashed, and solid lines
are the original continuum flux density, the flux density absorbed
by the homogeneous IGM, and the expected flux density after the
absorption by minihalos and dwarf galaxies, respectively.}
\label{Fig.DA}
\end{figure}

The existence of an X-ray background and its intensity are crucial
in determining the observability of the signals. If there is no
X-ray background, then with a GRB at redshift 9.8 and with the pixel
resolution of $\Delta\nu_{\rm ch} = 1.38\,\MHz$, we can get $19$
bands, then $19$ values of $D_{\rm A}$. We plot the $D_{\rm A}$ as a
function of observed frequency in the upper panel of
Fig.\ref{Fig.DA}, and with the Gaussian noises of $S/N = 5$, the
expected spectrum after the absorption by minihalos and dwarf
galaxies $F_\nu$ is shown in the bottom panel of Fig.\ref{Fig.DA}.
The original continuum $F_c$ and the flux density absorbed by the
homogeneous IGM $F_{\rm IGM}$ are also shown for comparison. With
SKA, we could hopefully detect the non-zero $D_{\rm A}$s if $f_{\rm
X}=0$ or it is extremely small, which are signals from both
minihalos and dwarf galaxies. In addition, $D_{\rm A}$ decreases
statistically with the decreasing redshift, which is a clear
signature of the evolution of non-linear structures during the epoch
of reionization. In the case of $f_{\rm X} = 0.05$, we could
marginally observe 2 pixels on a spectrum from $131.5 \MHz$ to $158
\MHz$, one of which has a band width of $\sim 8 \MHz$ ($131.5 -
139.5 \MHz$) with the mean flux decrement of $D_{\rm A} \approx
0.017$, and the other has a band width of $\sim 18.5 \MHz$ ($139.5 -
158 \MHz$) with the mean flux decrement of $D_{\rm A} \approx
0.009$. However, if $f_{\rm X}$ takes our fiducial value $0.1$, then
even one band spanning the whole spectrum from the source to the
lower redshift limit is still not broad enough to make the
observation feasible. The situation gets worse for higher values of
$f_{\rm X}$. Thus, the 21 cm broadband observation against high
redshift GRBs can be a powerful probe of the presence and the
intensity of the early X-ray background and the thermal evolution
during the epoch of reionization.

\section{DISCUSSION}\label{discuss}
We have modeled in detail the gas density and velocity profiles,
ionization state, temperature profiles, and Ly$\alpha$ photon
production for both minihalos and dwarf galaxies during the early
stages of reionization when the IGM was still patchy. We also take
into account an early X-ray background, which could partially ionize
and heat the IGM, and suppress the formation of low mass minihalos.
Using the detailed model, we investigate the spin temperature of
neutral hydrogen at different radii and the optical depth profiles
of 21 cm absorption lines, for various impact parameters, halo
masses, and redshifts.

We find that the Ly$\alpha$ background and the Ly$\alpha$ photons
from recombinations in the slightly ionized IGM couple the spin
temperature to the kinetic temperature outside minihalos or HII
regions of dwarf galaxies, and the coupling is already strong at
redshift 10. The Ly$\alpha$ background photons are blocked at
surfaces of halos, and the Ly$\alpha$ photons from recombinations
are negligible inside low-mass halos, though it dominates over the
collisional coupling in high-mass halos in which the gas is
partially ionized by collisions. The collisions also couple the spin
temperature to the gas kinetic temperature effectively at the
center, but this coupling decreases and the spin temperature
decreases with increasing radius.

The infalling velocity of gas around the minihalos/dwarf galaxies
plays a very important role in determining the profiles of the 21 cm
lines, double-peaked horn-like profiles are produced for a vast
range of parameters. The line profile of a dwarf galaxy also depends
on the size of its HII region, and hence its radiative properties.
The horn-like profile disappears for galaxies which are large, or
with top-heavy IMF so that the HII region expands beyond the turning
point of gas velocity.
The optical depth of a dwarf galaxy is lower for lines of sight
penetrating through its HII region, while a sufficiently large HII
region will lead to an optical depth trough rather than an
absorption line.

With the line number density based on halo mass function and a
physically motivated criterion for star formation, we generate
synthetic spectra of 21 cm forest by Monte Carlo procedure, and
calculate the cumulative distribution of equivalent width of the
absorption lines. Most of these lines have equivalent widths around
$0.03 \sim 0.3\,\kHz$ for an X-ray background intensity parameter of
$f_{\rm X}=0.1$, and the number of strong signals with large
equivalent widths decreases significantly with increasing $f_{\rm
X}$. We then study the observability of these signals. For high
resolution (1 KHz) observations, 
the GRB radio afterglows are too dim to be used as the background,
but absorption lines should be easily detected for a high redshift
quasar. It is exciting to know that the
Pan-STARRS\footnote{http://pan-starrs.ifa.hawaii.edu/public/home.html}
(the Panoramic Survey Telescope And Rapid Response System) is being
developed which will be able to detect quasars up to redshift $\sim
7$ and aims to find $\sim 20 - 50$ quasars at $z\sim 7$. For
broadband observations, it is also possible to detect the
absorptions against GRB radio afterglows if there is no X-ray
background. Setting a lower redshift cut-off at $z_{\rm lim} = 8$,
we find that the optimal redshift for the GRB is $z_{\rm GRB}\sim
9.8$. With a sensitivity of SKA, a signal-to-noise ratio of 5, and a
reasonable integration time, we could get measurements of mean flux
decrement $D_{\rm A}$ for 19 bands along the line of sight, each
with a channel width of $1.38\,\MHz$. In this way we could detect
not only the signals from both minihalos and dwarf galaxies, but
also their evolution during the epoch of reionization. However, the
detectability of 21 cm signals is very sensitive to the presence of
an early X-ray background. If an early X-ray background existed but
was not strong, taking $f_{\rm X}=0.05$, we could marginally observe
2 pixels along the line of sight towards a GRB which is located at
the optimal redshift of 9.8. Nonetheless, for the value of $f_{\rm
X}=0.1$ or higher, the signal will be impossible to be detected.
Therefore, the 21 cm absorption could be a powerful probe of the
presence/intensity of the X-ray background and the thermal history
in the early universe. However, we note that it is difficult to find
a radio source before the IGM has been significantly ionized or
heated, especially for a very bright high redshift quasar.

\citet{Furlanetto02} studied the 21 cm forest signals of minihalos
and early galaxies. Here, we have re-investigated this problem with
different and more detailed modeling of various properties of these
nonlinear objects, an early X-ray background, as well as the
Ly$\alpha$ background during the epoch of reionization. We found
stronger absorption signals from both minihalos and dwarf galaxies
for an early X-ray background not higher than the level today. Given
the many different model parameters adopted, this is not unexpected.
There are many differences in the details of modeling between the
two papers, but the main difference seems to be the IGM temperature.
\citet{Furlanetto02} have assumed a heated IGM with a simple form of
its evolution with redshift, which is already heated up to
$1000\,\K$ at $z=10$. This lies between our cases of $f_{\rm X} = 1$
and $f_{\rm X} = 5$. But in our fiducial model,
the gas temperature in the IGM is about $35\, \K$ at the same
redshift for $f_{\rm X}=0.1$. This is appropriate for the early
stages of stellar reionization, when the percolation has not
occurred yet. In \citet{Furlanetto02}, the gas structure in the
infalling region around minihalos/dwarf galaxies is modeled with a
self-similar solution of secondary infall found by
\citet{Bertschinger85}, while we have used the gas infall model
developed by \citet{Barkana04} which is based on the extended
Press-Schechter model and spherical collapse. The Bertschinger's
solution has a power law density profile of $\rho \propto
r^{-2.25}$, which is much steeper than the Barkana's prediction. Our
density and peculiar velocity structure of the infalling gas, with
the lower spin temperature outside the minihalos/HII regions,
produce a higher optical depth and a horn-like profile, which was
not found in \citet{Furlanetto02}.
For the dwarf galaxies,
\citet{Furlanetto02} considered protogalactic disks, whereas we assume here
a spherical symmetry in the gas density distribution since the
earliest galaxies are not likely to have large angular momentum.

In addition to the early X-ray background, an obvious uncertainty in
our model is the production of soft X-rays by dwarf galaxies in the
early universe. Especially with the normal IMF model-A and
relatively high metallicity, the amount of soft X-rays emitted after
a star burst is very uncertain \citep{Schaerer03}. However, this
amount of soft X-rays produced by stellar sources is always
negligible as compared to the background X-rays even if $f_{\rm
X}=0.05$. Another uncertainty comes from the gas density in
minihalos and dwarf galaxies. We have neglected the possible change
of density profile after the star formation. The gas density profile
could be modified by the expansion of the HII region when the
ionizing front changes nature from R-type to D-type. There are also
uncertainties in the star formation history in the dwarf galaxy. We
have assumed that the feedback effects quench subsequent star
formation (c.f. \citealt{Omukai99}), so that a single star burst is
produced. We have also assumed a smooth gas density distribution,
neglecting cool dense gas clumps from which the first stars are
likely to form. Thus we may have overestimated $R_{\rm HII}$, and
the optical depth could be slightly higher if they are accounted
for. However, the gas fraction in minihalos or dwarf galaxies could
also be lower than the cosmic mean value \citep{Naoz09}, then the
optical depth will be lower. Also, the results depend on the
assumption of $f_{\rm esc}$, which is taken to be 0.07, in agreement
with current observations.


As illustrated in Fig.\ref{Fig.depth_a} and Fig.\ref{Fig.depth_zM},
minihalos and dwarf galaxies exhibit distinct optical depth profiles
mainly due to the different ionization state and coupling physics, so they
are potentially distinguishable. Although we may not be able to
resolve the line profiles with the upcoming and planned instruments,
it is encouraging to be able to distinguish their features in a statistical
way, given that dwarf galaxies span a different halo population from
minihalos which cannot host stars. We reserve this investigation to
future works.


\section{ACKNOWLEDGMENTS}
We deeply appreciate the insight of the referee and the constructive
comments. We thank P. Dayal, S. Salvadori, W. Xu and B. Yue for
helpful discussions; we are grateful to R. Barkana who provided his
infall code. This work was supported in part by a scholarship from
China Scholarship council, by a research training fellowship from
SISSA astrophysics sector, by the NSFC grants 10373001, 10525314,
10533010, and 10773001, by the CAS grant KJCX3-SYW-N2, and by the
973 program No. 2007CB8125401.

{}
\end{document}